\documentclass[aps,amsmath,amssymb,preprint, twocolumn, 10pt]{revtex4-1}
\usepackage{amsmath, latexsym, color, graphicx, tabularx, booktabs, enumerate}
\usepackage{multirow}
\usepackage{dcolumn}
\usepackage{bm}
\usepackage{hyperref}
\usepackage{cleveref}
\usepackage{epstopdf}
\usepackage{titlesec}
\usepackage[toc,page]{appendix}
\titlespacing*{\section}{0pt}{1.1\baselineskip}{\baselineskip}
\date{\today}

\usepackage[normalem]{ulem} 
\begin{document}

\author{Niraj K. Nepal$^{1\dagger}$}
\author{Santosh Adhikari$^1$}
\author{Jefferson E. Bates$^2$}
\author{Adrienn Ruzsinszky$^1$}

\affiliation{[1] Department of Physics, Temple University, Philadelphia, Pennsylvania 19122, United States}
\affiliation{[2] A.R. Smith Department of Chemistry and Fermentation Sciences, Appalachian State University, Boone, North Carolina 28607, United States}

\title{Treating different bonding situations: Revisiting Au-Cu alloys using the random phase approximation}

\begin{abstract}
{\noindent The ground state equilibrium properties of copper-gold alloys have been explored with the state of art random phase approximation (RPA). Our estimated lattice constants agree with the experiment within a mean absolute percentage error (MAPE) of 1.4 percent. Semi-local functionals such as the generalized gradient
		approximation (GGA) of Perdew, Burke, and Ernzerhof (PBE) and strongly constrained and appropriately normed (SCAN) fail to provide accurate bulk moduli, which indicate their inability to describe the system in a stretched or compressed state with respect to the equilibrium geometry. We find that the non-locality present in RPA is able to describe the transition between two delocalized electron densities (bulk elemental constituents to crystallized alloys), as required to provide accurate formation energies. Based on our results, we conclude that it is difficult to find a universal density functional which can give accurate results for a wide range of properties of inter-metallic alloys. However, RPA can capture different bonding situations, often better than other density functionals. It gives accurate results for a wide range of ground state properties for the alloys, generated from metals with completely filled d-shells.}
\end{abstract}

\maketitle

\section{Introduction}
Inter-metallic alloys, the mixture of 2 or more metals in a solid form, manifest a defined stoichiometry and an ordered crystal structure \cite{O60}. The brittleness and high melting point with various electronic and magnetic properties of these solid compounds make them significantly useful in industrial applications. The heat of formation or formation energy of an alloy is the difference between the total binding energy of the system and its pure constituents, and its accurate prediction is extremely important in alloy theory. It governs the stability of the alloys with different compositions at different temperatures and pressures \cite{HDHGK73, O60}.\\

Density functional theory (DFT) is a robust electronic structure method, applicable across numerous fields of science. Various forms of the approximation to the exchange-correlation (XC-) energy, a key but unknown quantity in DFT, afford different levels of accuracy and computational efficiency. These different forms constitute the different rungs of Perdew's Jacob's ladder \cite{PS01}. The local density approximation (LDA)  \cite{KS65} and generalized gradient approximations (GGA) such as PBE  \cite{PBE96}, AM05 by Armiento and Mattsson  \cite{AM-05}, and PBE revised for solids (PBEsol) \cite{PRCVSCZB08}, though being highly accurate and efficient in many cases, often fail to provide an accurate description of the properties which require either self-interaction correction, at least a fraction of exact exchange, or an adequate description of many-body correlation  \cite{ZKP98}. Meta-GGAs such as TPSS \cite{TPSS03} by Tao, Perdew, Staroverov, and Scuseria, the made-simple (MS) family  \cite{SXR12, SHXBSP12}, and SCAN  \cite{SRP15} provide significant improvement over LDA and GGAs, with the inclusion of kinetic energy density ($\tau$(\textbf{r})) as an ingredient in addition to the density (n(\textbf{r})) and its gradient ($\nabla$n(\textbf{r})). However, regarding the formation energy of inter-metallic alloys, semi-local functionals have a mixed performance  \cite{IW18, ZKW14,OWZ98}. To estimate an accurate formation energy, a functional should not only provide reliable energetics of an alloy but also simultaneously of its constituent elements. However, most semi-local functionals fail in that regard, leading to inaccurate predictions of the formation energy  \cite{TLRKNV16}.\\

At high temperature, copper and gold form a continuous solid solution while they crystallize to form Au-Cu super-lattices at lower temperature \cite{ASM-87}. The Au-Cu systems, AuCu$_3$ (or Au$_{0.25}$Cu$_{0.75}$), AuCu (or Au$_{0.5}$Cu$_{0.5}$), and CuAu$_3$ (or Au$_{0.75}$Cu$_{0.25}$), are the paradigms of inter-metallic alloys, and have been extensively studied with semi-local as well as non-local DFT methods  \cite{ZKW14, TLRKNV16, OWZ98}. Experimentally, the fully ordered AuCu$_3$ and CuAu$_3$ stabilize in the L1$_2$ phase while the ordered AuCu prefers the L1$_0$ phase at T = 0 K  \cite{O60}. In addition to the distorted face-centered cubic (FCC) phase for AuCu (L1$_0$), we also have explored its FCC phase. Previous work clearly established that LDA could not predict the ground state of CuAu$_3$ as the L1$_2$ phase and also significantly underestimates the formation energies of all the Au-Cu alloys \cite{OWZ98}. Similar to LDA, PBE also predicts CuAu$_2$ as the stable phase and CuAu$_3$ as the unstable one  \cite{ZKW14}. On the other hand, the screened hybrid XC-functional HSE06 (simply HSE by Heyd, Scuseria, and Ernzerhof) \cite{HSE03, HSE03-er} developed by mixing non-local exact exchange with semi-local exchange-correlation, is able to provide accurate geometries and formation energies of these compounds  \cite{ZKW14}. Moreover, HSE is designed to be nonlocal at short range, while the exact exchange is screened at long range. SCAN was demonstrated to capture medium-range weak interactions, while HSE is known to reduce the delocalization error in semilocal functionals. Inspired by these facts, we aim to climb even higher on the rungs of Jacob's ladder to gain more understanding about any competition between weak interactions and free electron-like bonding in Cu-Au systems \cite{SGKMK13}. \\

In this work, we have revisited the Au-Cu alloy systems using the random phase approximation (RPA)  \cite{BP52, BP53, LP80}. RPA is the simplest approximation within an adiabatic-connection fluctuation-dissipation theorem (ACFDT) formalism \cite{LP75, LP77}. It combines the non-local one electron self-interaction free exact exchange (EXX) energy with the non-local correlation energy (E$^{RPA}_c$)  \cite{EBF12, RRJ12}. Most importantly, it can provide accurate results for systems involving weak interactions such as van der Waals (vdW) interactions  \cite{HK08,SHS10,LHG10}, as well as ionic  \cite{NRB18, SGKMK13} and covalent interactions  \cite{HSK10,PL13,SGKMK13,OT13}. The total energy in the ACFDT-DFT framework can be expressed as,
\begin{equation}
\resizebox{0.4\hsize}{!}{$%
E = E_{EXX} + E_C^{RPA}
$}
\end{equation}
where, E$_{EXX}$ is the Hartree-Fock (HF) total energy evaluated non-self-consistently using self-consistently obtained Kohn-Sham DFT orbitals. E$_C^{RPA}$ is the RPA correlation energy, which can be obtained using the interacting density-density response function ($\chi$) which is related to the non-interacting response function ($\chi_0$) via a Dyson-like equation  \cite{EBF12, RRJ12, F01}. The
RPA correlation energy naturally incorporates long-range dispersion and is non-perturbative. For this work, the systems are heavy coinage metals which are largely influenced by dispersion interactions \cite{BEH09, MT78, FW12}. Furthermore, the systems have zero band gap. Due to its non-perturbative nature, RPA can be safely applied to zero-gap systems without divergence  \cite{CHG14}. In both aspects, the application of RPA to these systems is justified.\\ 

For the sake of comparison, we also have assessed semi-local functionals such as PBE, PBEsol, the revised TPSS (revTPSS) \cite{SMCRHKKP11} and SCAN along with RPA. The rest of the paper is organized as follows. Computational details are provided in section II, followed by results in section III. We will present our conclusions in section IV.
\section{computational details}
All DFT calculations were carried out using a projected augmented wave (PAW)  \cite{B94} method, as implemented in GPAW  \cite{gpaw1,gpaw2,ase} and VASP  \cite{VASP}. We utilized VASP to perform semi-local calculations whereas RPA calculations were carried out using GPAW. Moreover, semi-local calculations were performed self-consistently while RPA calculations were carried out using a non-self-consistent approach. We used a plane-wave cutoff of 600 eV and Brillouin zone sampling of 20$\times$20$\times$20 Gamma centered k-mesh to avoid the convergence test for semi-local DFT calculations. Ground state PBE calculations were performed as an input for the RPA calculations. Separate convergence tests for EXX and the RPA correlation energies were carried out to determine the plane-wave cutoffs and k-mesh sampling with less than 2 meV relative error ({\color{blue} Appendix~\ref{tab:append}}). We used a maximum cutoff of 350 eV to compute the response function. All other parameters and procedure of the RPA calculations were kept similar to that of Ref.~\onlinecite{NRB18}, except skipping the gamma point (\textbf{q} $=$ 0) to avoid the possible divergent contribution from metals as discussed in Ref. ~\onlinecite{HSK10}.\\

We calculated the zero-point vibrational energy (ZPVE) to estimate the thermal contribution to the formation energy. PBEsol  \cite{PRCVSCZB08} calculations were done with a 2$\times$2$\times$2 supercell (32 atoms) using VASP  \cite{KF96, KJ99} and PHONOPY  \cite{TT15}. The estimated thermal corrections are less than or equal to 1 meV/atom, consistent with previous results  \cite{ZKW14}. Relativistic effects are included at the scalar level for each atom within the PAW potentials provided in VASP and GPAW. We have performed calculations for 7 volume points near the experimental equilibrium volume and fit the Birch-Murnaghan equation of state  \cite{B71} to evaluate the equilibrium properties. We have used the experimental structures from Ref.~\onlinecite{ASM-87} and varied the lattice constants isotropically to generate structures with different volumes. In order to compute the cohesive energies, atomic energies were computed. We have performed spin polarized semi-local DFT calculations with VASP using a plane-wave cutoff of 600 eV and 23 $\times$ 24 $\times$ 25 $\AA^3$ simulation cell to avoid any interactions of an isolated atom with its periodic images. Separate convergence tests for atomic energies were performed with GPAW for both EXX and RPA ({\color{blue} Appendix~\ref{tab:append}}).  

\section{Results}
\subsection{Lattice constant} 
The equilibrium lattice constants of the ordered Au-Cu alloys are presented in Table~\ref{latt:const}. As expected, PBE overestimates the lattice constants and PBEsol yields reasonable lattice constants of the coinage metals such as Au and Cu \cite{SGKMK13}. The failure of PBE to estimate accurate lattice constants is related to the poor descriptions of correlation effects between completely filled d-shells in coinage metals \cite{SGKMK13}. RPA lattice constants are also overestimated which can be decreased by including the Pauli repulsion in SOSEX or adding a kernel correction to RPA \cite{BLR16, NRB18, GMHSK09}, thereby improving the short-range correlations necessary to describe the systems with more filled d-shells. On the other hand, SCAN along with revTPSS show good performance in the prediction of equilibrium lattice constant. With the inclusion of kinetic energy density, both SCAN and revTPSS can distinguish different bonding regions relevant to lattice constants, and this becomes more effective as more d bands are filled in the transition metal \cite{SGKMK13}. Overall, all methods show a reasonable agreement with the experiment for the lattice constant with mean absolute percentage error (MAPE) less than 1.4\%. 
\begin{table}[h!]
	\caption{Lattice constants ($\AA$). The experimental lattice constants are taken from Reference~\onlinecite{ASM-87}. Among the DFT functionals utilized, PBEsol shows the best performance in predicting the equilibrium lattice constant.}
	\resizebox{0.9 \linewidth}{!}{%
	\begin{tabular}{l|r|r|r|r|r|r}
		\hline \hline
		& \multicolumn{1}{l|}{PBE} & PBEsol & revTPSS & \multicolumn{1}{l|}{SCAN} & \multicolumn{1}{l|}{RPA} & \multicolumn{1}{l}{Experiment} \\ \hline
		Cu & 3.64 & 3.57 & 3.57 & 3.57 &  3.63 & 3.62 \\ 
		Au & 4.16 & 4.08 & 4.08 & 4.09 &  4.15 & 4.08 \\ 
		AuCu$_3$ & 3.78 & 3.72 & 3.72 & 3.72   & 3.78 & 3.75 \\ 
		AuCu (FCC) & 3.92 & 3.85 & 3.85 & 3.85   & 3.92 & 3.87 \\ 
		AuCu (P4/mmm) & 2.84 & 2.79 & 2.79 & 2.79   & 2.83 & 2.80 \\
		CuAu$_3$ & 4.05 & 3.97 & 3.98 & 3.98  & 4.05 & 3.95 \\ \hline
		MAE ($\AA$) & 0.051 & 0.022 & 0.023 & 0.026  & 0.046 & -- \\ 
		MAPE (\%) & 1.37 & 0.59 & 0.64 & 0.7 &   1.21 & -- \\
		\hline\hline
	\end{tabular}
}
	\label{latt:const}
\end{table}
\subsection{Bulk Moduli}
The bulk modulus measures the curvature of an energy-volume relation, and its accurate prediction indicates the ability of a DFT approximation to describe the system in a non-equilibrium state with respect to the equilibrium state. We calculated the bulk moduli of various Au-Cu systems and tabulated them in Table~\ref{tab:bm}. PBE underestimates the bulk modulus, while PBEsol and revTPSS predict accurate bulk moduli for gold. However, the performance of these functionals worsens on increasing the concentration of copper. On the other hand, SCAN provides an improvement for the bulk modulus of copper, thereby improving the bulk moduli of all alloys compared to PBEsol and revTPSS. The inability of semi-local functionals to give an accurate prediction of bulk moduli indicates their inability to describe the compressed or stretched electron densities with respect to the equilibrium ground state electron densities. One can see that the overall bulk moduli predicted by PBE for all the alloys are close to the experimental value of copper while those of other semi-local functionals are closer to gold. 
On the other hand, RPA provides accurate bulk moduli for both alloys and those of the constituent bulk elemental systems. This indicates that, contrary to the lattice constant, the description of short-range correlation is not crucial for the prediction of bulk modulus. 

\begin{table}[h!]
	\caption{Bulk Modulus (GPa). Experimental bulk moduli are computed using $\frac{(C_{11} + 2C_{12})}{3}$ (cubic lattice), where C$_{ij}$ is the elastic moduli. Overall, RPA predicts the bulk moduli in close agreement with the experiment.}
	\resizebox{\columnwidth}{!}{%
	\begin{tabular}{l|r|r|r|r|r|r|l}
		\hline \hline
		& \multicolumn{1}{l|}{PBE} & PBEsol & revTPSS & \multicolumn{1}{l|}{SCAN} & \multicolumn{1}{l|}{HSE \cite{JLKVLTI14}}  & \multicolumn{1}{l|}{RPA} & Experiment \\ \hline
		Cu & 137.89 & 164.50 & 170.14 & 157.91 & 133.8 & 144.74 &  \multicolumn{1}{r}{143.6  \cite{H58}} \\ 
		Au & 139.03 & 174.34 & 176.01 & 166.95 & 146.6  & 176.71 & 177.6  \cite{C18}, 180.53  \cite{H58} \\ 
		AuCu$_3$ & 139.49 & 168.05 & 171.09 & 164.84 &   & 155.25 & \multicolumn{1}{r}{151.83  \cite{OM71}} \\ 
	    AuCu (FCC) & 139.99 & 171.02 & 173.12 & 169.47 &   & 163.77 & \multicolumn{1}{r}{162.97  \cite{T13}} \\ 
	    AuCu (P4/mmm) & 138.75 & 169.56 & 171.90 & 166.16 &   & 159.09  & \multicolumn{1}{r}{--} \\ 
		CuAu$_3$ & 139.03 & 171.64 & 173.25 & 165.74 &   & 162.15 & 166.33 \cite{S65}  \\ \hline
		MAE (GPa) & 21.71 & 11.04 & 13.19 & 9.31  & -- & 2.42 & \\ 
		MAPE (\%) & 13.02 & 7.20 & 8.65 & 5.93 &  -- & 1.50 & \\
		\hline \hline
	
	\end{tabular}
}
	\label{tab:bm}
\end{table}

\subsection{Formation Energy}
\begin{figure*}[h!]
 	\caption{Formation energy (eV) with respect to volume ($\AA^3$). The positive formation energy refers to instability of alloys with respect to isolated bulk constituents, while negative formation energy refers to stability. Experimental formation energies are taken form References~\onlinecite{O60, HDHGK73}. The experimental volumes are indicated by `\#' in Figures.}
 	\includegraphics[scale=0.3]{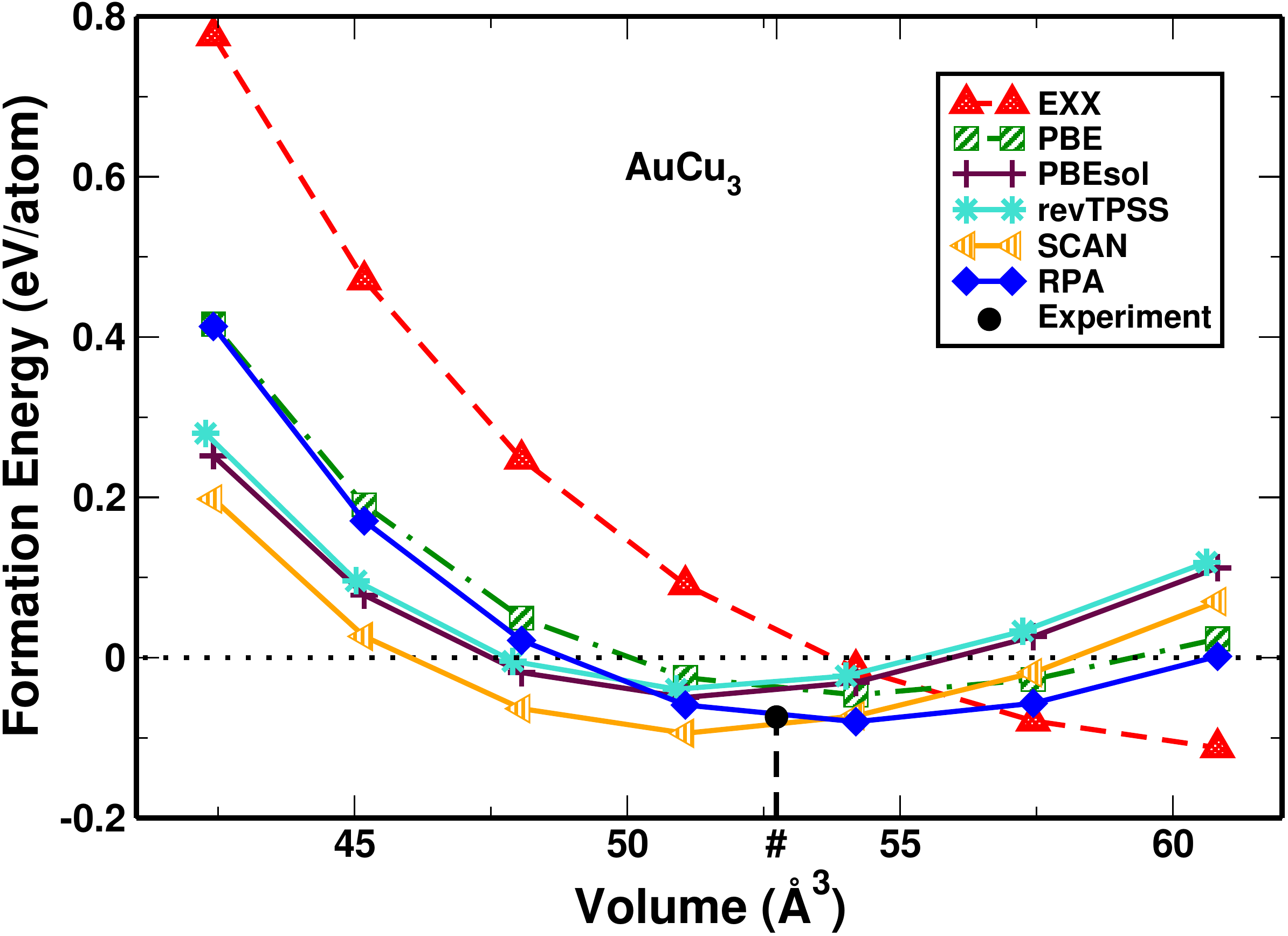}
 	\includegraphics[scale=0.3]{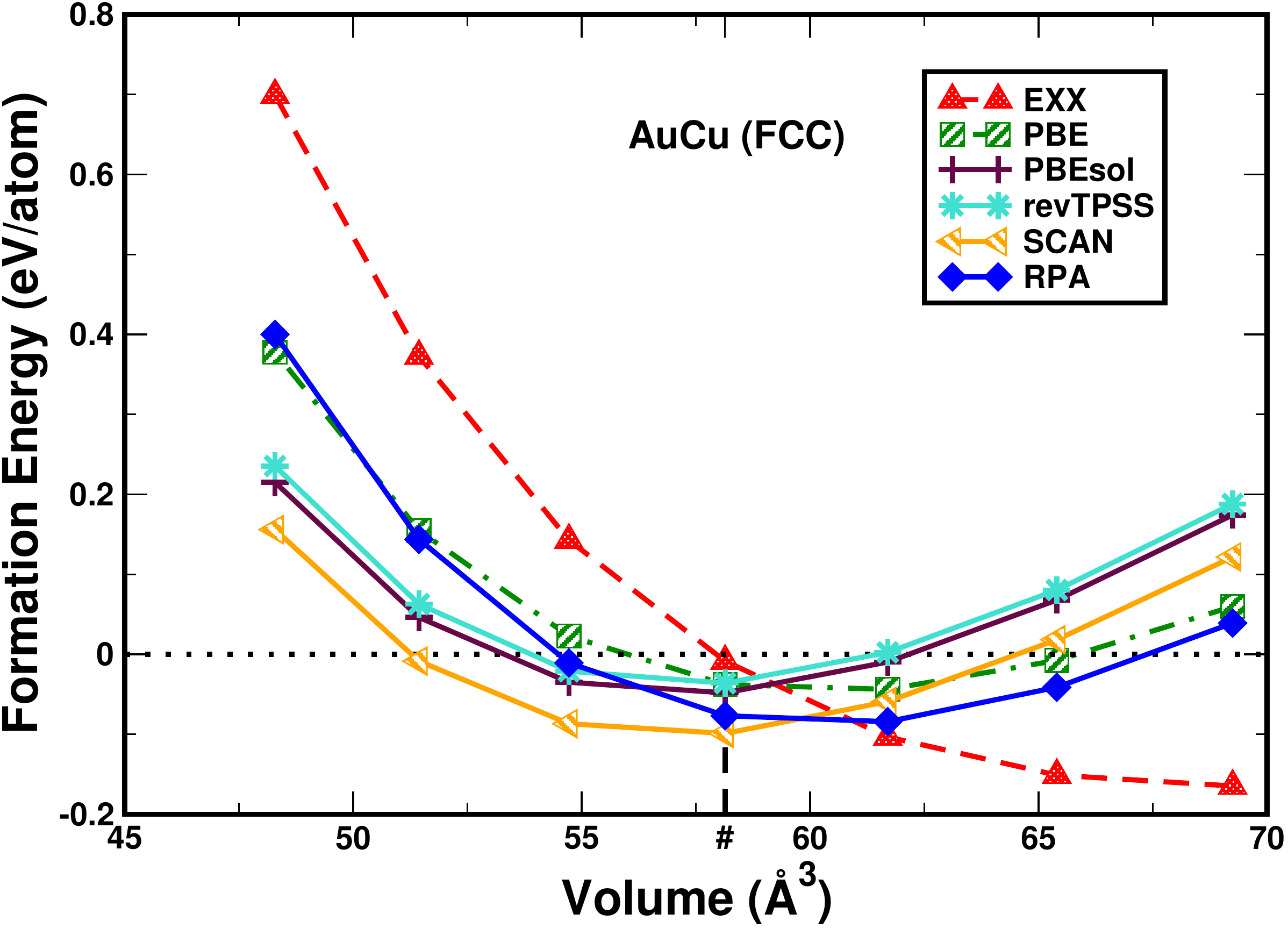}
 	\vspace{1cm}
 	\includegraphics[scale=0.3]{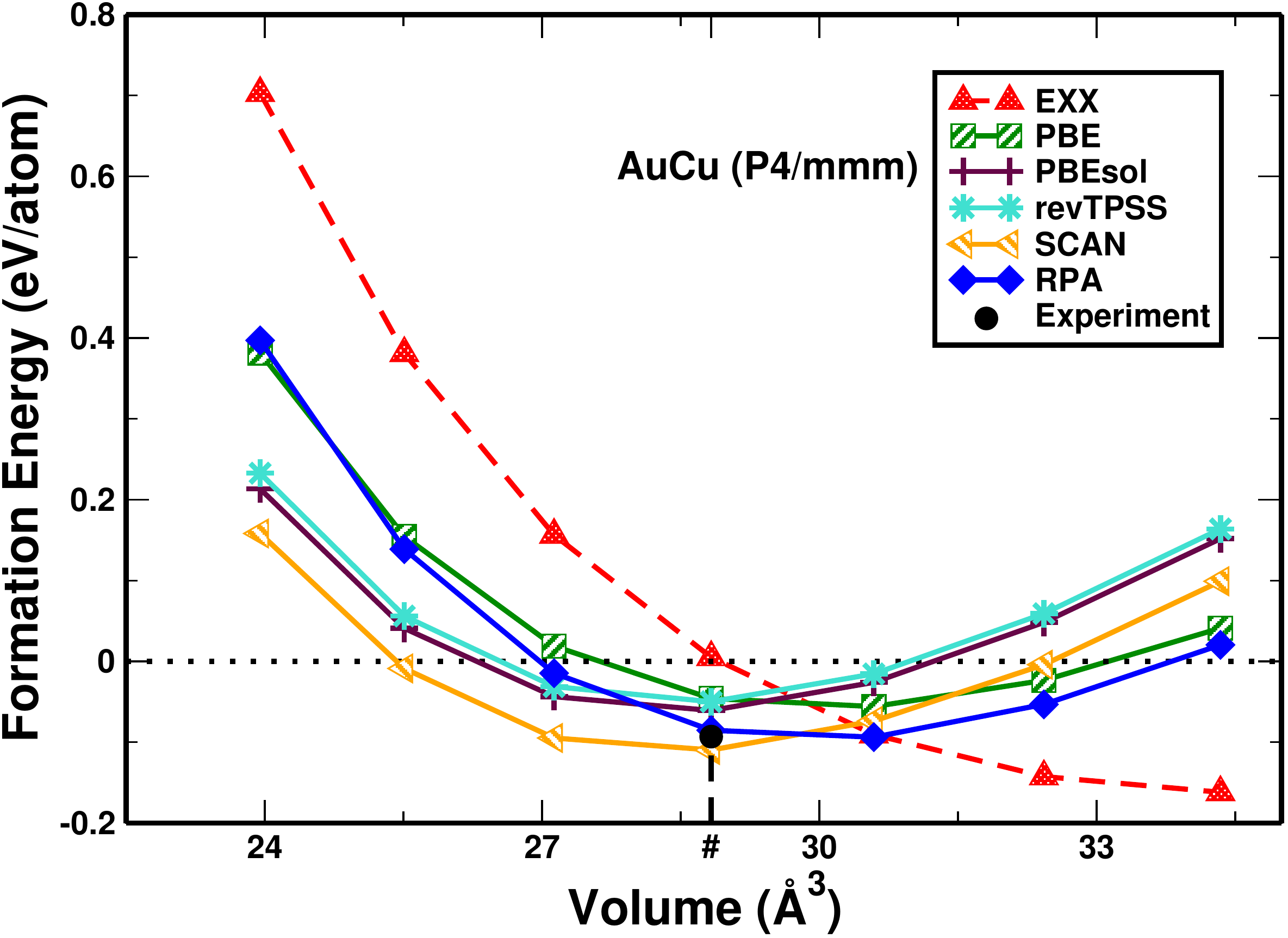}
 	\includegraphics[scale=0.3]{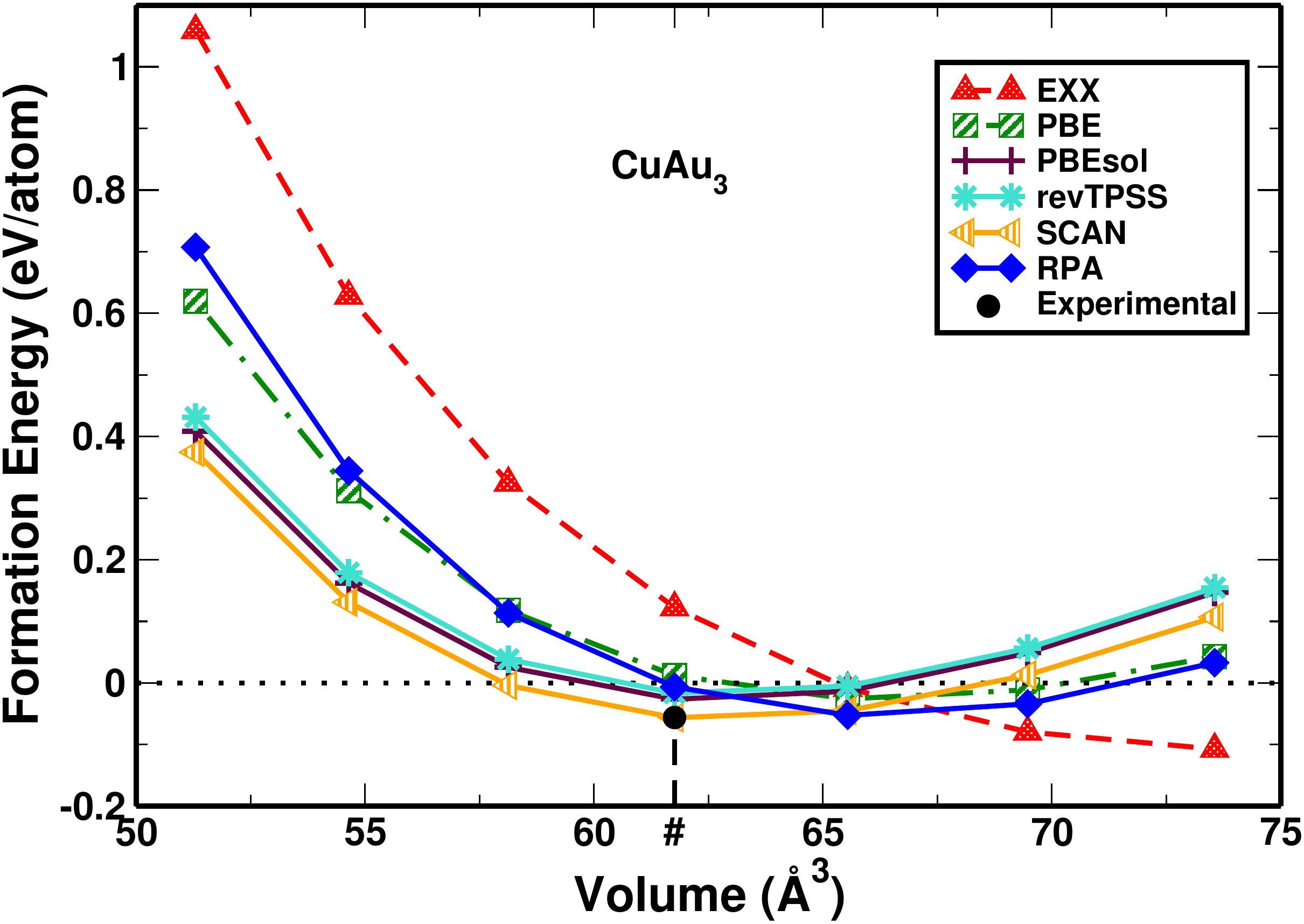}
 	\label{fig:EvsV}
 \end{figure*}

The formation energy of a Au-Cu alloy can be obtained using,
\begin{equation}
\resizebox{0.9\hsize}{!}{$%
E_f(Au_xCu_{1-x}) = E_{coh}(Au_xCu_{1-x}) - xE_{coh}(Au) - (1-x)E_{coh}(Cu)
$}
\label{fm}
\end{equation}
where E$_f$(Au$_x$Cu$_{1-x}$), E$_{coh}$(Au$_x$Cu$_{1-x}$), E$_{coh}$(Au), and E$_{coh}$(Cu) are the formation energy of system, cohesive energy of the whole system, cohesive energy of gold, and the cohesive energy of copper per atom respectively and $x$ is the fraction of gold atoms in the alloys. We have computed the formation energy of a given alloy as a function of crystal volume, Figure~\ref{fig:EvsV}. Experimental volumes are indicated by a `\#', whereas the formation energies are represented by black dots. Positive formation energies imply instability of the alloy while negative formation energies imply stability with respect to their elemental bulk constituents.\\ 

Without any correlation, EXX predicts destabilized systems within the range of volumes considered. Contrary to EXX, PBE stabilizes the systems with the presence of both exchange and correlation energies (Figure~\ref{fig:EvsV}). Formation energies obtained from PBE agree with RPA values either at the highly compressed state or at the highly stretched state. However, they start to deviate from RPA and even from the experimental value as the equilibrium geometry is approached. On the contrary, SCAN slightly overbinds the formation energies near the equilibrium geometries, and the overestimation gets larger as we deviate from the equilibrium. The formation energy vs volume calculated with PBEsol and revTPSS behave similarly to that of SCAN, however, they are slightly shifted upward along the direction of positive formation energy. Despite the fact that RPA overestimates the lattice constant, it accurately predicts the curvature and minima of the equation of state for these alloys.\\

\begin{figure*}[h!]
	\caption{Convex hull: Formation energy as a function of gold composition (x). PBE along with PBEsol and revTPSS largely underestimates the formation energy, while SCAN performs poorly on Cu-rich alloy, but improves the result as the concentration of gold increases. Formation energies predicted by non-local HSE and RPA are in close agreement with the experiment.}
	\includegraphics[scale = 0.4]{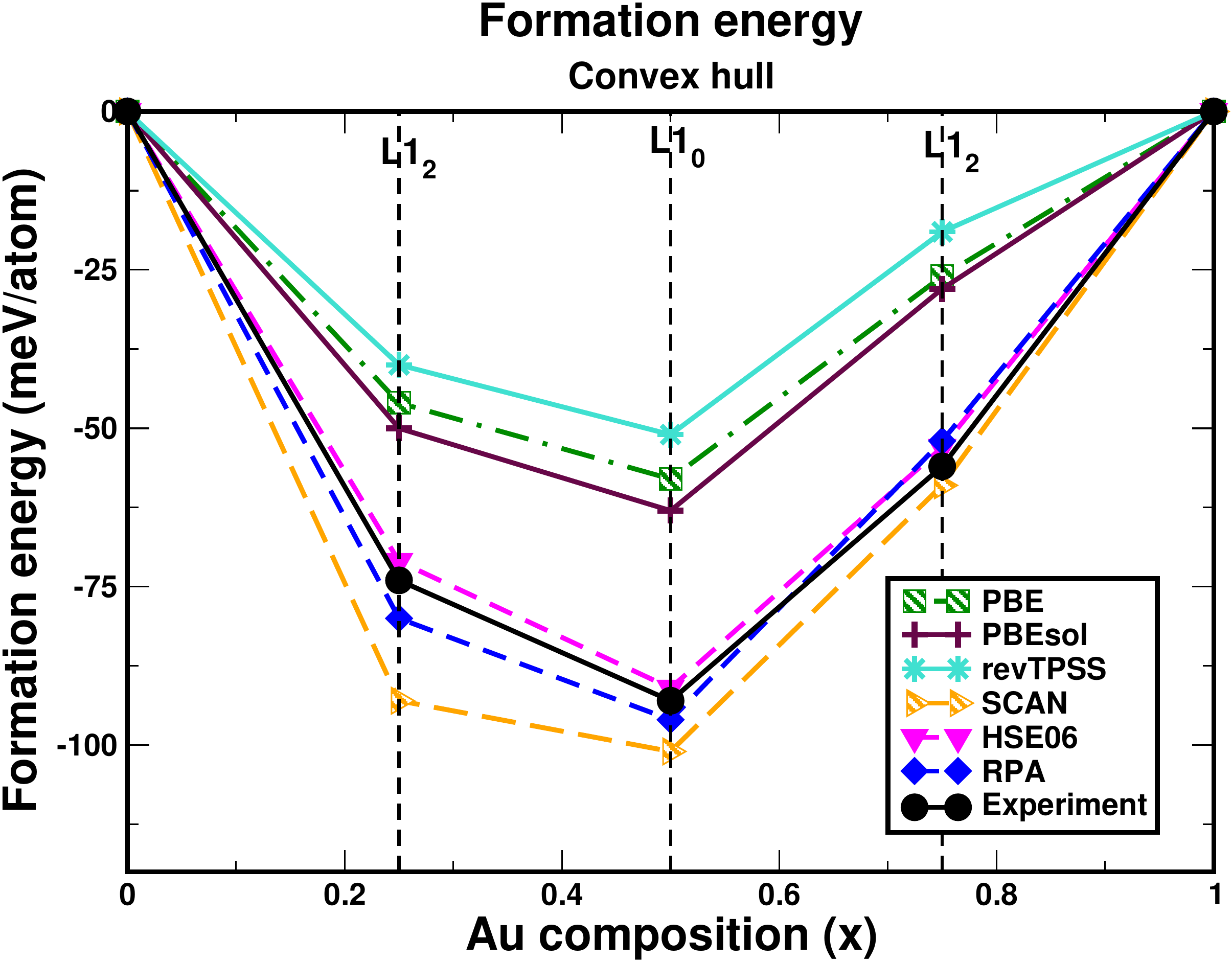}
	\label{fig:convex}
\end{figure*}

The heats of formation of the Au-Cu system at equilibrium are presented in Table~\ref{tab:FE}, whereas the stability of the alloys are represented by a convex hull as in Figure~\ref{fig:convex}. As in the earlier studies \cite{ZKW14, TLRKNV16}, PBE severely underestimates the formation energies. On restoring the second-order gradient expansion for the exchange over a wide range of densities at the GGA level, PBEsol slightly improves the results, but at the meta-GGA level revTPSS \cite{PRCCS09} worsens it. By satisfying more exact constraints and including more appropriate norms \cite{SRP15}, SCAN shows a considerable improvement over revTPSS. However, it still performs poorly for copper-rich alloys, while the error decreases on increasing the concentration of gold, as more filled 5d shells in Au are involved. In contrast, non-local density functionals such as hybrid HSE and RPA consistently predict accurate formation energies of these alloys. In comparison, self-consistent HSE outperforms non-self-consistent RPA by only a little, but there can be room for improvement when RPA is also evaluated self-consistently \cite{NCG14}.\\

In the present work, both alloys and the constituent metals are in the solid phase and hence possess delocalized electron densities. The reliable prediction of formation energies requires an accurate description of the transition from delocalized electron densities of constituent metals to delocalized electron densities of inter-metallic alloys. All semi-local functionals included herein fail to describe such a transition, while the non-locality present in HSE and RPA is able to effectively detect such changes. However, the transition from localized (atoms) to delocalized (solid metals and alloys) is not straightforward even for non-local functionals, as evident from the cohesive energies presented in Table ~\ref{tab:coh}. The better density functional for formation energies is the worst for cohesive energies. The hybrid HSE seriously underestimates the cohesive energies of Au-Cu system with a mean absolute error (MAE) of nearly 0.7 eV/atom. Surprisingly, semi-local functionals perform much better than HSE in the order of PBE $<$ revTPSS $<$ PBEsol $<$ SCAN with decreasing MAE and MAPE. RPA, on the other hand, provides reasonable cohesive energies for Cu-rich compounds, while it worsens on increasing the concentration of gold in the alloys. As the number of filled d bands increases,  the short-range correlation becomes more crucial in describing the interactions within transition metal atoms as well as their alloys. Restoring the exchange-correlation kernel within the RPA can improve the cohesive energies of transition metals up to 0.3 to 0.4 eV \cite{OT13, JOBT15}.
\vspace*{0.3cm}

The performance of various density functionals on Au-Cu alloys clearly depends on their ability to describe the less-delocalized ``3d" electron density of copper as well as the more-delocalized ``5d" electron density of gold. SCAN along with PBEsol and revTPSS can effectively describe the 5d bands of gold, thereby giving sensibly accurate lattice constants, bulk moduli, and cohesive energies for Au-rich alloys. However, RPA has the opposite trend that it can provide an accurate prediction for Cu-rich alloys, but falls short when describing Au-rich alloys. On the contrary, HSE fails to provide accurate bulk moduli and cohesive energies for both copper and gold, giving too low cohesive energies for both Au- and Cu- rich alloys. With that in mind, one can argue that the density functionals that can separately describe the constituents, can ultimately describe the weakly bonded alloys.

\begin{table}[h!]
	\caption{The heat of formation (eV /atom). Note that the results for semi-local functionals and HSE are obtained self-consistently, while the RPA results are obtained non-self-consistently using PBE orbitals. Experimental results taken from Reference~\onlinecite{O60} are obtained at 320 K, whereas Reference~\onlinecite{HDHGK73} corresponds to 298.15 K; CuAu$_3$: The experimental structure taken in Reference~\onlinecite{O60} is not fully ordered structure. The heat of formation for fully ordered CuAu$_3$ is estimated using cubic interpolation of composition (x)-Gibbs energy ($\Delta$G)-entropy ($\Delta$S) of formation data taken from Reference~\onlinecite{HDHGK73}.}
	\resizebox{\columnwidth}{!}{%
	\begin{tabular}{l|r|r|r|r|r|r|r}
		\hline
		\hline
		& \multicolumn{1}{l|}{PBE} & PBEsol & revTPSS & \multicolumn{1}{l|}{SCAN} & \multicolumn{1}{l|}{HSE \cite{ZKW14}} & \multicolumn{1}{l|}{RPA} & \multicolumn{1}{l}{Experiment \cite{O60}} \\ 
		\hline
		AuCu3 & -0.046 & -0.050 & -0.040 & -0.093 & -0.071 & -0.080  & -0.074 , -0.075 \cite{HDHGK73} \\ 
		AuCu (FCC) & -0.047 & -0.050 & -0.037 & -0.101 &   & -0.088  & \multicolumn{1}{l}{} \\ 
		AuCu (P4/mmm) & -0.058 & -0.062 & -0.051 & -0.111 & -0.091 & -0.096 & -0.093  \\
		CuAu3 & -0.026 & -0.028 & -0.019 & -0.059 & -0.053 & -0.052 & -0.039, -0.056 \cite{HDHGK73} \\ \hline
		\hline
	\end{tabular}
}
	\label{tab:FE}
\end{table}
 
\begin{table}[h!]
	\caption{Cohesive energy per atom (eV/atom). The cohesive energies for HSE are obtained using References~\onlinecite{JLKVLTI14, ZKW14}, while the experimental cohesive energies are obtained using References~\onlinecite{O60, KMM76} using Eq.~\ref{fm}. Among the functionals used, SCAN predicts most accurate cohesive energy as compared to the experiment.}
	\resizebox{\linewidth}{!}{%
		\begin{tabular}{l|r|r|r|r|r|r|r}
			\hline
			\hline
			& \multicolumn{1}{l|}{PBE} & PBEsol & revTPSS & \multicolumn{1}{l|}{SCAN} & \multicolumn{1}{l|}{HSE} & \multicolumn{1}{l|}{RPA} & \multicolumn{1}{l}{Experiment} \\ 
			\hline
			Cu & 3.484 & 4.030 & 4.057 & 3.886 & 3.060 \cite{JLKVLTI14}  & 3.350  & 3.490 \cite{KMM76}  \\
			Au & 3.035 & 3.720 & 3.627 & 3.555 & 2.880 \cite{JLKVLTI14} & 3.395 & 3.810 \cite{KMM76} \\
			AuCu3 & 3.419 & 4.003 & 3.990 & 3.896 & 3.086 & 3.441 & 3.644 \\ 
			AuCu (P4/mmm) & 3.317 & 3.937 & 3.893 & 3.828 & 3.111  & 3.469 & 3.743  \\
			CuAu3 & 3.175 & 3.826 & 3.753 & 3.695 & 2.978 & 3.436 &  3.786 \\ \hline
			MAE (eV/atom) & 0.409 & 0.245 & 0.256 &  0.216  & 0.672 & 0.276 & -- \\ 
			MAPE (\%) & 10.84 & 6.78 & 7.08 & 5.93 & 18.05   & 7.41  & -- \\
			\hline
			\hline
		\end{tabular}
	}
	\label{tab:coh}
\end{table}
\vspace{0.2cm}
\section{Conclusions}
We have explored copper-gold alloys with various levels of approximations within density functional theory, including the state-of-art random phase approximation. It is difficult to find a universal functional which can describe all of the bonding situations. Semi-local functionals can reasonably describe the transition between localized and delocalized electron densities, as manifested in the cohesive energies. On the other hand, the non-locality present in HSE and RPA can distinguish the transition between two delocalized electron densities, as seen from the formation energies of the alloys. Moreover, the performance of these functionals in describing the weakly bonded Au-Cu system depends on their potential to separately describe less-delocalized and more-delocalized electron densities of copper and gold respectively. Based on our results, we can conclude that RPA predicts accurate values for diverse properties of binary alloys, generated from metals with completely filled d-shells. It has an accuracy of semi-local functionals at the challenging situations, while consistently providing reliable results where semi-local functionals break down.      
\section{Acknowledgements}
N.K.N. and A.R. acknowledge support by the National Science Foundation under Grant No. DMR-1553022. JEB was supported by the A.R. Smith Department of Chemistry and Fermentation Sciences. Computational support was provided by Temple University's HPC resources and thus was supported in part by the National Science Foundation through major research instrumentation grant number 1625061 and by the US Army Research Laboratory under contract number W911NF-16-2-0189. N.K.N., JEB, and A.R. designed the project. N.K.N. performed the calculations, and wrote the first draft. S.A. performed some of the calculations. All authors contributed references, discussions, and revisions. 
\clearpage
$^\dagger$\textbf{\textit{niraj.nepal@temple.edu}}
\appendix
\section{Appendix: Convergence tests}
\begin{table}[h!]
	\renewcommand\thetable{A}
	\caption{Convergence parameters for RPA calculations; Bulk calculations: Energies for bulk calculations are converged within 1-2 meV; Atomic calculations: EXX energies are converged within 1-2 meV for both energy cutoff and cell volume. However for RPA, convergence was achieved within 1-5 meV for energy cutoff, while it is 15-20 meV for cell volume. We used PAW pseudopotentials such that the valence electron configuration includes d$^{10}$S$^1$ electrons for both copper and gold.}
	\resizebox{1.6\columnwidth}{!}{%
		\begin{tabular}{|l|r|l|r|l|l|l|l|l|}
			\hline
			& \multicolumn{1}{l}{} & \multicolumn{1}{l}{Bulk calculations} & \multicolumn{1}{l}{} & \multicolumn{1}{l|}{} & \multicolumn{1}{l}{Atomic calculations} & \multicolumn{1}{l}{} & \multicolumn{1}{l}{}& \\ \hline
			& \multicolumn{1}{l}{EXX} &  & \multicolumn{1}{l}{RPA} &  & \multicolumn{1}{l}{EXX} &  & \multicolumn{1}{l}{RPA} &\\ \hline
			& \multicolumn{1}{l|}{Energy cutoff (eV)} & K-mesh & \multicolumn{1}{l|}{Energy cutoff (eV)} & K-mesh & Energy cutoff (eV) & Cell volume ($\AA^3$) & Energy cutoff (eV) & Cell volume ($\AA^3$)  \\ \hline
			Cu & 1000 & 16$\times$16$\times$16 & 800 & 16$\times$16$\times$16 & \multicolumn{1}{r|}{1000} & 10$\times$11$\times$12 & \multicolumn{1}{r|}{600} & 8$\times$9$\times$10 \\ \hline
			Au & 1000 & 16$\times$16$\times$16 & 800 & 16$\times$16$\times$16 & \multicolumn{1}{r|}{900} & 9$\times$10$\times$11 & \multicolumn{1}{r|}{500} & 8$\times$9$\times$10 \\ \hline
			AuCu3 & 1000 & 16$\times$16$\times$16 & 800 & 16$\times$16$\times$16 &  &  &  &\\ \hline
			AuCu & 1000 & 15$\times$15$\times$15 & 800 & 16$\times$16$\times$16 &  &  &  &\\ \hline
			CuAu3 & 800 & 15$\times$15$\times$15 & 800 & 16$\times$16$\times$16 &  &  &  &\\ \hline
		\end{tabular}
	}
    \label{tab:append}
\end{table}


\begin{thebibliography}{61}%
	\makeatletter
	\providecommand \@ifxundefined [1]{%
		\@ifx{#1\undefined}
	}%
	\providecommand \@ifnum [1]{%
		\ifnum #1\expandafter \@firstoftwo
		\else \expandafter \@secondoftwo
		\fi
	}%
	\providecommand \@ifx [1]{%
		\ifx #1\expandafter \@firstoftwo
		\else \expandafter \@secondoftwo
		\fi
	}%
	\providecommand \natexlab [1]{#1}%
	\providecommand \enquote  [1]{``#1''}%
	\providecommand \bibnamefont  [1]{#1}%
	\providecommand \bibfnamefont [1]{#1}%
	\providecommand \citenamefont [1]{#1}%
	\providecommand \href@noop [0]{\@secondoftwo}%
	\providecommand \href [0]{\begingroup \@sanitize@url \@href}%
	\providecommand \@href[1]{\@@startlink{#1}\@@href}%
	\providecommand \@@href[1]{\endgroup#1\@@endlink}%
	\providecommand \@sanitize@url [0]{\catcode `\\12\catcode `\$12\catcode
		`\&12\catcode `\#12\catcode `\^12\catcode `\_12\catcode `\%12\relax}%
	\providecommand \@@startlink[1]{}%
	\providecommand \@@endlink[0]{}%
	\providecommand \url  [0]{\begingroup\@sanitize@url \@url }%
	\providecommand \@url [1]{\endgroup\@href {#1}{\urlprefix }}%
	\providecommand \urlprefix  [0]{URL }%
	\providecommand \Eprint [0]{\href }%
	\providecommand \doibase [0]{http://dx.doi.org/}%
	\providecommand \selectlanguage [0]{\@gobble}%
	\providecommand \bibinfo  [0]{\@secondoftwo}%
	\providecommand \bibfield  [0]{\@secondoftwo}%
	\providecommand \translation [1]{[#1]}%
	\providecommand \BibitemOpen [0]{}%
	\providecommand \bibitemStop [0]{}%
	\providecommand \bibitemNoStop [0]{.\EOS\space}%
	\providecommand \EOS [0]{\spacefactor3000\relax}%
	\providecommand \BibitemShut  [1]{\csname bibitem#1\endcsname}%
	\let\auto@bib@innerbib\@empty
	\bibitem [{\citenamefont {Orr}(1960)}]{O60}%
	\BibitemOpen
	\bibfield  {author} {\bibinfo {author} {\bibfnamefont {R.}~\bibnamefont
			{Orr}},\ }\href@noop {} {\bibfield  {journal} {\bibinfo  {journal} {Acta
				Metall.}\ }\textbf {\bibinfo {volume} {8}},\ \bibinfo {pages} {489} (\bibinfo
		{year} {1960})}\BibitemShut {NoStop}%
	\bibitem [{\citenamefont {Hultgren}\ \emph {et~al.}(1973)\citenamefont
		{Hultgren}, \citenamefont {Desai}, \citenamefont {Hawkins}, \citenamefont
		{Gleiser},\ and\ \citenamefont {Kelley}}]{HDHGK73}%
	\BibitemOpen
	\bibfield  {author} {\bibinfo {author} {\bibfnamefont {R.}~\bibnamefont
			{Hultgren}}, \bibinfo {author} {\bibfnamefont {P.~D.}\ \bibnamefont {Desai}},
		\bibinfo {author} {\bibfnamefont {D.~T.}\ \bibnamefont {Hawkins}}, \bibinfo
		{author} {\bibfnamefont {M.}~\bibnamefont {Gleiser}}, \ and\ \bibinfo
		{author} {\bibfnamefont {K.~K.}\ \bibnamefont {Kelley}},\ }\href@noop {}
	{\emph {\bibinfo {title} {Selected values of the thermodynamic properties of
				binary alloys}}},\ \bibinfo {type} {Tech. Rep.}\ (\bibinfo  {institution}
	{National Standard Reference Data System},\ \bibinfo {year}
	{1973})\BibitemShut {NoStop}%
	\bibitem [{\citenamefont {Perdew}\ and\ \citenamefont {Schmidt}(2001)}]{PS01}%
	\BibitemOpen
	\bibfield  {author} {\bibinfo {author} {\bibfnamefont {J.~P.}\ \bibnamefont
			{Perdew}}\ and\ \bibinfo {author} {\bibfnamefont {K.}~\bibnamefont
			{Schmidt}},\ }in\ \href@noop {} {\emph {\bibinfo {booktitle} {Density
				Functional Theory and Its Applications to Materials}}},\ Vol.\ \bibinfo
	{volume} {577},\ \bibinfo {editor} {edited by\ \bibinfo {editor}
		{\bibfnamefont {V.}~\bibnamefont {{Van Doren}}}, \bibinfo {editor}
		{\bibfnamefont {C.}~\bibnamefont {{Van Alsenoy}}}, \ and\ \bibinfo {editor}
		{\bibfnamefont {P.}~\bibnamefont {Geerlings}}}\ (\bibinfo  {publisher} {AIP
		Conference Proceedings, Melville, N.Y.},\ \bibinfo {year} {2001})\ pp.\
	\bibinfo {pages} {1--20}\BibitemShut {NoStop}%
	\bibitem [{\citenamefont {Kohn}\ and\ \citenamefont {Sham}(1965)}]{KS65}%
	\BibitemOpen
	\bibfield  {author} {\bibinfo {author} {\bibfnamefont {W.}~\bibnamefont
			{Kohn}}\ and\ \bibinfo {author} {\bibfnamefont {L.~J.}\ \bibnamefont
			{Sham}},\ }\href@noop {} {\bibfield  {journal} {\bibinfo  {journal} {Phys.
				Rev.}\ }\textbf {\bibinfo {volume} {140}},\ \bibinfo {pages} {A1133}
		(\bibinfo {year} {1965})}\BibitemShut {NoStop}%
	\bibitem [{\citenamefont {Perdew}\ \emph {et~al.}(1996)\citenamefont {Perdew},
		\citenamefont {Burke},\ and\ \citenamefont {Ernzerhof}}]{PBE96}%
	\BibitemOpen
	\bibfield  {author} {\bibinfo {author} {\bibfnamefont {J.~P.}\ \bibnamefont
			{Perdew}}, \bibinfo {author} {\bibfnamefont {K.}~\bibnamefont {Burke}}, \
		and\ \bibinfo {author} {\bibfnamefont {M.}~\bibnamefont {Ernzerhof}},\
	}\href@noop {} {\bibfield  {journal} {\bibinfo  {journal} {Phys. Rev. Lett.}\
		}\textbf {\bibinfo {volume} {77}},\ \bibinfo {pages} {3865} (\bibinfo {year}
		{1996})}\BibitemShut {NoStop}%
	\bibitem [{\citenamefont {Armiento}\ and\ \citenamefont
		{Mattsson}(2005)}]{AM-05}%
	\BibitemOpen
	\bibfield  {author} {\bibinfo {author} {\bibfnamefont {R.}~\bibnamefont
			{Armiento}}\ and\ \bibinfo {author} {\bibfnamefont {A.~E.}\ \bibnamefont
			{Mattsson}},\ }\href@noop {} {\bibfield  {journal} {\bibinfo  {journal}
			{Phys. Rev. B}\ }\textbf {\bibinfo {volume} {72}},\ \bibinfo {pages} {085108}
		(\bibinfo {year} {2005})}\BibitemShut {NoStop}%
	\bibitem [{\citenamefont {Perdew}\ \emph {et~al.}(2008)\citenamefont {Perdew},
		\citenamefont {Ruzsinszky}, \citenamefont {Csonka}, \citenamefont {Vydrov},
		\citenamefont {Scuseria}, \citenamefont {Constantin}, \citenamefont {Zhou},\
		and\ \citenamefont {Burke}}]{PRCVSCZB08}%
	\BibitemOpen
	\bibfield  {author} {\bibinfo {author} {\bibfnamefont {J.~P.}\ \bibnamefont
			{Perdew}}, \bibinfo {author} {\bibfnamefont {A.}~\bibnamefont {Ruzsinszky}},
		\bibinfo {author} {\bibfnamefont {G.~I.}\ \bibnamefont {Csonka}}, \bibinfo
		{author} {\bibfnamefont {O.~A.}\ \bibnamefont {Vydrov}}, \bibinfo {author}
		{\bibfnamefont {G.~E.}\ \bibnamefont {Scuseria}}, \bibinfo {author}
		{\bibfnamefont {L.~A.}\ \bibnamefont {Constantin}}, \bibinfo {author}
		{\bibfnamefont {X.}~\bibnamefont {Zhou}}, \ and\ \bibinfo {author}
		{\bibfnamefont {K.}~\bibnamefont {Burke}},\ }\href@noop {} {\bibfield
		{journal} {\bibinfo  {journal} {Phys. Rev. Lett.}\ }\textbf {\bibinfo
			{volume} {100}},\ \bibinfo {pages} {136406} (\bibinfo {year}
		{2008})}\BibitemShut {NoStop}%
	\bibitem [{\citenamefont {Ziesche}\ \emph {et~al.}(1998)\citenamefont
		{Ziesche}, \citenamefont {Kurth},\ and\ \citenamefont {Perdew}}]{ZKP98}%
	\BibitemOpen
	\bibfield  {author} {\bibinfo {author} {\bibfnamefont {P.}~\bibnamefont
			{Ziesche}}, \bibinfo {author} {\bibfnamefont {S.}~\bibnamefont {Kurth}}, \
		and\ \bibinfo {author} {\bibfnamefont {J.~P.}\ \bibnamefont {Perdew}},\
	}\href@noop {} {\bibfield  {journal} {\bibinfo  {journal} {Comput. Mater.
				Sci.}\ }\textbf {\bibinfo {volume} {11}},\ \bibinfo {pages} {122} (\bibinfo
		{year} {1998})}\BibitemShut {NoStop}%
	\bibitem [{\citenamefont {Tao}\ \emph {et~al.}(2003)\citenamefont {Tao},
		\citenamefont {Perdew}, \citenamefont {Staroverov},\ and\ \citenamefont
		{Scuseria}}]{TPSS03}%
	\BibitemOpen
	\bibfield  {author} {\bibinfo {author} {\bibfnamefont {J.}~\bibnamefont
			{Tao}}, \bibinfo {author} {\bibfnamefont {J.~P.}\ \bibnamefont {Perdew}},
		\bibinfo {author} {\bibfnamefont {V.~N.}\ \bibnamefont {Staroverov}}, \ and\
		\bibinfo {author} {\bibfnamefont {G.~E.}\ \bibnamefont {Scuseria}},\
	}\href@noop {} {\bibfield  {journal} {\bibinfo  {journal} {Phys. Rev. Lett.}\
		}\textbf {\bibinfo {volume} {91}},\ \bibinfo {pages} {146401} (\bibinfo
		{year} {2003})}\BibitemShut {NoStop}%
	\bibitem [{\citenamefont {Sun}\ \emph {et~al.}(2012)\citenamefont {Sun},
		\citenamefont {Xiao},\ and\ \citenamefont {Ruzsinszky}}]{SXR12}%
	\BibitemOpen
	\bibfield  {author} {\bibinfo {author} {\bibfnamefont {J.}~\bibnamefont
			{Sun}}, \bibinfo {author} {\bibfnamefont {B.}~\bibnamefont {Xiao}}, \ and\
		\bibinfo {author} {\bibfnamefont {A.}~\bibnamefont {Ruzsinszky}},\ }\href
	{\doibase 10.1063/1.4742312} {\bibfield  {journal} {\bibinfo  {journal} {J.
				Chem. Phys.}\ }\textbf {\bibinfo {volume} {137}},\ \bibinfo {pages} {051101}
		(\bibinfo {year} {2012})}\BibitemShut {NoStop}%
	\bibitem [{\citenamefont {Sun}\ \emph {et~al.}(2013)\citenamefont {Sun},
		\citenamefont {Haunschild}, \citenamefont {Xiao}, \citenamefont {Bulik},
		\citenamefont {Scuseria},\ and\ \citenamefont {Perdew}}]{SHXBSP12}%
	\BibitemOpen
	\bibfield  {author} {\bibinfo {author} {\bibfnamefont {J.}~\bibnamefont
			{Sun}}, \bibinfo {author} {\bibfnamefont {R.}~\bibnamefont {Haunschild}},
		\bibinfo {author} {\bibfnamefont {B.}~\bibnamefont {Xiao}}, \bibinfo {author}
		{\bibfnamefont {I.~W.}\ \bibnamefont {Bulik}}, \bibinfo {author}
		{\bibfnamefont {G.~E.}\ \bibnamefont {Scuseria}}, \ and\ \bibinfo {author}
		{\bibfnamefont {J.~P.}\ \bibnamefont {Perdew}},\ }\href@noop {} {\bibfield
		{journal} {\bibinfo  {journal} {J. Chem. Phys.}\ }\textbf {\bibinfo {volume}
			{138}},\ \bibinfo {pages} {044113} (\bibinfo {year} {2013})}\BibitemShut
	{NoStop}%
	\bibitem [{\citenamefont {Sun}\ \emph {et~al.}(2015)\citenamefont {Sun},
		\citenamefont {Ruzsinszky},\ and\ \citenamefont {Perdew}}]{SRP15}%
	\BibitemOpen
	\bibfield  {author} {\bibinfo {author} {\bibfnamefont {J.}~\bibnamefont
			{Sun}}, \bibinfo {author} {\bibfnamefont {A.}~\bibnamefont {Ruzsinszky}}, \
		and\ \bibinfo {author} {\bibfnamefont {J.~P.}\ \bibnamefont {Perdew}},\
	}\href@noop {} {\bibfield  {journal} {\bibinfo  {journal} {Phys. Rev. Lett.}\
		}\textbf {\bibinfo {volume} {115}},\ \bibinfo {pages} {036402} (\bibinfo
		{year} {2015})}\BibitemShut {NoStop}%
	\bibitem [{\citenamefont {Isaacs}\ and\ \citenamefont
		{Wolverton}(2018)}]{IW18}%
	\BibitemOpen
	\bibfield  {author} {\bibinfo {author} {\bibfnamefont {E.~B.}\ \bibnamefont
			{Isaacs}}\ and\ \bibinfo {author} {\bibfnamefont {C.}~\bibnamefont
			{Wolverton}},\ }\href@noop {} {\bibfield  {journal} {\bibinfo  {journal}
			{Phys. Rev. Mater.}\ }\textbf {\bibinfo {volume} {2}},\ \bibinfo {pages}
		{063801} (\bibinfo {year} {2018})}\BibitemShut {NoStop}%
	\bibitem [{\citenamefont {Zhang}\ \emph {et~al.}(2014)\citenamefont {Zhang},
		\citenamefont {Kresse},\ and\ \citenamefont {Wolverton}}]{ZKW14}%
	\BibitemOpen
	\bibfield  {author} {\bibinfo {author} {\bibfnamefont {Y.}~\bibnamefont
			{Zhang}}, \bibinfo {author} {\bibfnamefont {G.}~\bibnamefont {Kresse}}, \
		and\ \bibinfo {author} {\bibfnamefont {C.}~\bibnamefont {Wolverton}},\
	}\href@noop {} {\bibfield  {journal} {\bibinfo  {journal} {Phys. Rev. Lett.}\
		}\textbf {\bibinfo {volume} {112}},\ \bibinfo {pages} {075502} (\bibinfo
		{year} {2014})}\BibitemShut {NoStop}%
	\bibitem [{\citenamefont {Ozoli{\c{n}}{\v{s}}}\ \emph
		{et~al.}(1998)\citenamefont {Ozoli{\c{n}}{\v{s}}}, \citenamefont
		{Wolverton},\ and\ \citenamefont {Zunger}}]{OWZ98}%
	\BibitemOpen
	\bibfield  {author} {\bibinfo {author} {\bibfnamefont {V.}~\bibnamefont
			{Ozoli{\c{n}}{\v{s}}}}, \bibinfo {author} {\bibfnamefont {C.}~\bibnamefont
			{Wolverton}}, \ and\ \bibinfo {author} {\bibfnamefont {A.}~\bibnamefont
			{Zunger}},\ }\href@noop {} {\bibfield  {journal} {\bibinfo  {journal} {Phys.
				Rev. B}\ }\textbf {\bibinfo {volume} {57}},\ \bibinfo {pages} {6427}
		(\bibinfo {year} {1998})}\BibitemShut {NoStop}%
	\bibitem [{\citenamefont {Tian}\ \emph {et~al.}(2016)\citenamefont {Tian},
		\citenamefont {Lev{\"a}m{\"a}ki}, \citenamefont {Ropo}, \citenamefont
		{Kokko}, \citenamefont {Nagy},\ and\ \citenamefont {Vitos}}]{TLRKNV16}%
	\BibitemOpen
	\bibfield  {author} {\bibinfo {author} {\bibfnamefont {L.-Y.}\ \bibnamefont
			{Tian}}, \bibinfo {author} {\bibfnamefont {H.}~\bibnamefont
			{Lev{\"a}m{\"a}ki}}, \bibinfo {author} {\bibfnamefont {M.}~\bibnamefont
			{Ropo}}, \bibinfo {author} {\bibfnamefont {K.}~\bibnamefont {Kokko}},
		\bibinfo {author} {\bibfnamefont {{\'A}.}~\bibnamefont {Nagy}}, \ and\
		\bibinfo {author} {\bibfnamefont {L.}~\bibnamefont {Vitos}},\ }\href@noop {}
	{\bibfield  {journal} {\bibinfo  {journal} {Phys. Rev. Lett.}\ }\textbf
		{\bibinfo {volume} {117}},\ \bibinfo {pages} {066401} (\bibinfo {year}
		{2016})}\BibitemShut {NoStop}%
	\bibitem [{\citenamefont {Okamoto}\ \emph {et~al.}(1987)\citenamefont
		{Okamoto}, \citenamefont {Chakrabarti}, \citenamefont {Laughlin},\ and\
		\citenamefont {Massalski}}]{ASM-87}%
	\BibitemOpen
	\bibfield  {author} {\bibinfo {author} {\bibfnamefont {H.}~\bibnamefont
			{Okamoto}}, \bibinfo {author} {\bibfnamefont {D.~J.}\ \bibnamefont
			{Chakrabarti}}, \bibinfo {author} {\bibfnamefont {D.~E.}\ \bibnamefont
			{Laughlin}}, \ and\ \bibinfo {author} {\bibfnamefont {T.~B.}\ \bibnamefont
			{Massalski}},\ }\href {\doibase 10.1007/BF02893155} {\bibfield  {journal}
		{\bibinfo  {journal} {Journal of Phase Equilibria}\ }\textbf {\bibinfo
			{volume} {8}},\ \bibinfo {pages} {454} (\bibinfo {year} {1987})}\BibitemShut
	{NoStop}%
	\bibitem [{\citenamefont {Heyd}\ \emph {et~al.}(2003)\citenamefont {Heyd},
		\citenamefont {Scuseria},\ and\ \citenamefont {Ernzerhof}}]{HSE03}%
	\BibitemOpen
	\bibfield  {author} {\bibinfo {author} {\bibfnamefont {J.}~\bibnamefont
			{Heyd}}, \bibinfo {author} {\bibfnamefont {G.~E.}\ \bibnamefont {Scuseria}},
		\ and\ \bibinfo {author} {\bibfnamefont {M.}~\bibnamefont {Ernzerhof}},\
	}\href@noop {} {\bibfield  {journal} {\bibinfo  {journal} {J. chem. phys.}\
		}\textbf {\bibinfo {volume} {118}},\ \bibinfo {pages} {8207} (\bibinfo {year}
		{2003})}\BibitemShut {NoStop}%
	\bibitem [{\citenamefont {Heyd}\ \emph {et~al.}(2006)\citenamefont {Heyd},
		\citenamefont {Scuseria},\ and\ \citenamefont {Ernzerhof}}]{HSE03-er}%
	\BibitemOpen
	\bibfield  {author} {\bibinfo {author} {\bibfnamefont {J.}~\bibnamefont
			{Heyd}}, \bibinfo {author} {\bibfnamefont {G.~E.}\ \bibnamefont {Scuseria}},
		\ and\ \bibinfo {author} {\bibfnamefont {M.}~\bibnamefont {Ernzerhof}},\
	}\href@noop {} {\bibfield  {journal} {\bibinfo  {journal} {The Journal of
				Chemical Physics}\ }\textbf {\bibinfo {volume} {124}},\ \bibinfo {pages}
		{219906} (\bibinfo {year} {2006})}\BibitemShut {NoStop}%
	\bibitem [{\citenamefont {Schimka}\ \emph {et~al.}(2013)\citenamefont
		{Schimka}, \citenamefont {Gaudoin}, \citenamefont {Klime{\v{s}}},
		\citenamefont {Marsman},\ and\ \citenamefont {Kresse}}]{SGKMK13}%
	\BibitemOpen
	\bibfield  {author} {\bibinfo {author} {\bibfnamefont {L.}~\bibnamefont
			{Schimka}}, \bibinfo {author} {\bibfnamefont {R.}~\bibnamefont {Gaudoin}},
		\bibinfo {author} {\bibfnamefont {J.}~\bibnamefont {Klime{\v{s}}}}, \bibinfo
		{author} {\bibfnamefont {M.}~\bibnamefont {Marsman}}, \ and\ \bibinfo
		{author} {\bibfnamefont {G.}~\bibnamefont {Kresse}},\ }\href@noop {}
	{\bibfield  {journal} {\bibinfo  {journal} {Phys. Rev. B}\ }\textbf {\bibinfo
			{volume} {87}},\ \bibinfo {pages} {214102} (\bibinfo {year}
		{2013})}\BibitemShut {NoStop}%
	\bibitem [{\citenamefont {Bohm}\ and\ \citenamefont {Pines}(1952)}]{BP52}%
	\BibitemOpen
	\bibfield  {author} {\bibinfo {author} {\bibfnamefont {D.}~\bibnamefont
			{Bohm}}\ and\ \bibinfo {author} {\bibfnamefont {D.}~\bibnamefont {Pines}},\
	}\href@noop {} {\bibfield  {journal} {\bibinfo  {journal} {Phys. Rev.}\
		}\textbf {\bibinfo {volume} {85}},\ \bibinfo {pages} {338} (\bibinfo {year}
		{1952})}\BibitemShut {NoStop}%
	\bibitem [{\citenamefont {Bohm}\ and\ \citenamefont {Pines}(1953)}]{BP53}%
	\BibitemOpen
	\bibfield  {author} {\bibinfo {author} {\bibfnamefont {D.}~\bibnamefont
			{Bohm}}\ and\ \bibinfo {author} {\bibfnamefont {D.}~\bibnamefont {Pines}},\
	}\href {\doibase 10.1103/PhysRev.92.609} {\bibfield  {journal} {\bibinfo
			{journal} {Phys. Rev.}\ }\textbf {\bibinfo {volume} {92}},\ \bibinfo {pages}
		{609} (\bibinfo {year} {1953})}\BibitemShut {NoStop}%
	\bibitem [{\citenamefont {Langreth}\ and\ \citenamefont {Perdew}(1980)}]{LP80}%
	\BibitemOpen
	\bibfield  {author} {\bibinfo {author} {\bibfnamefont {D.~C.}\ \bibnamefont
			{Langreth}}\ and\ \bibinfo {author} {\bibfnamefont {J.~P.}\ \bibnamefont
			{Perdew}},\ }\href@noop {} {\bibfield  {journal} {\bibinfo  {journal} {Phys.
				Rev. B}\ }\textbf {\bibinfo {volume} {21}},\ \bibinfo {pages} {5469}
		(\bibinfo {year} {1980})}\BibitemShut {NoStop}%
	\bibitem [{\citenamefont {Langreth}\ and\ \citenamefont {Perdew}(1975)}]{LP75}%
	\BibitemOpen
	\bibfield  {author} {\bibinfo {author} {\bibfnamefont {D.~C.}\ \bibnamefont
			{Langreth}}\ and\ \bibinfo {author} {\bibfnamefont {J.~P.}\ \bibnamefont
			{Perdew}},\ }\href@noop {} {\bibfield  {journal} {\bibinfo  {journal} {Solid
				State Commun.}\ }\textbf {\bibinfo {volume} {17}},\ \bibinfo {pages} {1425}
		(\bibinfo {year} {1975})}\BibitemShut {NoStop}%
	\bibitem [{\citenamefont {Langreth}\ and\ \citenamefont {Perdew}(1977)}]{LP77}%
	\BibitemOpen
	\bibfield  {author} {\bibinfo {author} {\bibfnamefont {D.~C.}\ \bibnamefont
			{Langreth}}\ and\ \bibinfo {author} {\bibfnamefont {J.~P.}\ \bibnamefont
			{Perdew}},\ }\href@noop {} {\bibfield  {journal} {\bibinfo  {journal} {Phys.
				Rev. B}\ }\textbf {\bibinfo {volume} {15}},\ \bibinfo {pages} {2884}
		(\bibinfo {year} {1977})}\BibitemShut {NoStop}%
	\bibitem [{\citenamefont {Eshuis}\ \emph {et~al.}(2012)\citenamefont {Eshuis},
		\citenamefont {Bates},\ and\ \citenamefont {Furche}}]{EBF12}%
	\BibitemOpen
	\bibfield  {author} {\bibinfo {author} {\bibfnamefont {H.}~\bibnamefont
			{Eshuis}}, \bibinfo {author} {\bibfnamefont {J.~E.}\ \bibnamefont {Bates}}, \
		and\ \bibinfo {author} {\bibfnamefont {F.}~\bibnamefont {Furche}},\
	}\href@noop {} {\bibfield  {journal} {\bibinfo  {journal} {Theor. Chem.
				Acc.}\ }\textbf {\bibinfo {volume} {131}},\ \bibinfo {pages} {1084} (\bibinfo
		{year} {2012})}\BibitemShut {NoStop}%
	\bibitem [{\citenamefont {Ren}\ \emph {et~al.}(2012)\citenamefont {Ren},
		\citenamefont {Rinke}, \citenamefont {Joas},\ and\ \citenamefont
		{Scheffler}}]{RRJ12}%
	\BibitemOpen
	\bibfield  {author} {\bibinfo {author} {\bibfnamefont {X.}~\bibnamefont
			{Ren}}, \bibinfo {author} {\bibfnamefont {P.}~\bibnamefont {Rinke}}, \bibinfo
		{author} {\bibfnamefont {C.}~\bibnamefont {Joas}}, \ and\ \bibinfo {author}
		{\bibfnamefont {M.}~\bibnamefont {Scheffler}},\ }\href {\doibase
		10.1007/s10853-012-6570-4} {\bibfield  {journal} {\bibinfo  {journal} {J.
				Mater. Sci.}\ }\textbf {\bibinfo {volume} {47}},\ \bibinfo {pages} {7447}
		(\bibinfo {year} {2012})}\BibitemShut {NoStop}%
	\bibitem [{\citenamefont {Harl}\ and\ \citenamefont {Kresse}(2008)}]{HK08}%
	\BibitemOpen
	\bibfield  {author} {\bibinfo {author} {\bibfnamefont {J.}~\bibnamefont
			{Harl}}\ and\ \bibinfo {author} {\bibfnamefont {G.}~\bibnamefont {Kresse}},\
	}\href@noop {} {\bibfield  {journal} {\bibinfo  {journal} {Phys. Rev. B}\
		}\textbf {\bibinfo {volume} {77}},\ \bibinfo {pages} {045136} (\bibinfo
		{year} {2008})}\BibitemShut {NoStop}%
	\bibitem [{\citenamefont {Schimka}\ \emph {et~al.}(2010)\citenamefont
		{Schimka}, \citenamefont {Harl}, \citenamefont {Stroppa}, \citenamefont
		{Gr\"{u}neis}, \citenamefont {Marsman}, \citenamefont {Mittendorfer},\ and\
		\citenamefont {Kresse}}]{SHS10}%
	\BibitemOpen
	\bibfield  {author} {\bibinfo {author} {\bibfnamefont {L.}~\bibnamefont
			{Schimka}}, \bibinfo {author} {\bibfnamefont {J.}~\bibnamefont {Harl}},
		\bibinfo {author} {\bibfnamefont {A.}~\bibnamefont {Stroppa}}, \bibinfo
		{author} {\bibfnamefont {A.}~\bibnamefont {Gr\"{u}neis}}, \bibinfo {author}
		{\bibfnamefont {M.}~\bibnamefont {Marsman}}, \bibinfo {author} {\bibfnamefont
			{F.}~\bibnamefont {Mittendorfer}}, \ and\ \bibinfo {author} {\bibfnamefont
			{G.}~\bibnamefont {Kresse}},\ }\href {\doibase 10.1038/nmat2806} {\bibfield
		{journal} {\bibinfo  {journal} {Nat. Mater.}\ }\textbf {\bibinfo {volume}
			{9}},\ \bibinfo {pages} {741} (\bibinfo {year} {2010})}\BibitemShut {NoStop}%
	\bibitem [{\citenamefont {Leb{\`{e}}gue}\ \emph {et~al.}(2010)\citenamefont
		{Leb{\`{e}}gue}, \citenamefont {Harl}, \citenamefont {Gould}, \citenamefont
		{{\'{A}}ngy{\'{a}}n}, \citenamefont {Kresse},\ and\ \citenamefont
		{Dobson}}]{LHG10}%
	\BibitemOpen
	\bibfield  {author} {\bibinfo {author} {\bibfnamefont {S.}~\bibnamefont
			{Leb{\`{e}}gue}}, \bibinfo {author} {\bibfnamefont {J.}~\bibnamefont {Harl}},
		\bibinfo {author} {\bibfnamefont {T.}~\bibnamefont {Gould}}, \bibinfo
		{author} {\bibfnamefont {J.~G.}\ \bibnamefont {{\'{A}}ngy{\'{a}}n}}, \bibinfo
		{author} {\bibfnamefont {G.}~\bibnamefont {Kresse}}, \ and\ \bibinfo {author}
		{\bibfnamefont {J.~F.}\ \bibnamefont {Dobson}},\ }\href@noop {} {\bibfield
		{journal} {\bibinfo  {journal} {Phys. Rev. Lett.}\ }\textbf {\bibinfo
			{volume} {105}},\ \bibinfo {pages} {196401} (\bibinfo {year}
		{2010})}\BibitemShut {NoStop}%
	\bibitem [{\citenamefont {Nepal}\ \emph {et~al.}(2018)\citenamefont {Nepal},
		\citenamefont {Ruzsinszky},\ and\ \citenamefont {Bates}}]{NRB18}%
	\BibitemOpen
	\bibfield  {author} {\bibinfo {author} {\bibfnamefont {N.~K.}\ \bibnamefont
			{Nepal}}, \bibinfo {author} {\bibfnamefont {A.}~\bibnamefont {Ruzsinszky}}, \
		and\ \bibinfo {author} {\bibfnamefont {J.~E.}\ \bibnamefont {Bates}},\ }\href
	{\doibase 10.1103/PhysRevB.97.115140} {\bibfield  {journal} {\bibinfo
			{journal} {Phys. Rev. B}\ }\textbf {\bibinfo {volume} {97}},\ \bibinfo
		{pages} {115140} (\bibinfo {year} {2018})}\BibitemShut {NoStop}%
	\bibitem [{\citenamefont {Harl}\ \emph {et~al.}(2010)\citenamefont {Harl},
		\citenamefont {Schimka},\ and\ \citenamefont {Kresse}}]{HSK10}%
	\BibitemOpen
	\bibfield  {author} {\bibinfo {author} {\bibfnamefont {J.}~\bibnamefont
			{Harl}}, \bibinfo {author} {\bibfnamefont {L.}~\bibnamefont {Schimka}}, \
		and\ \bibinfo {author} {\bibfnamefont {G.}~\bibnamefont {Kresse}},\
	}\href@noop {} {\bibfield  {journal} {\bibinfo  {journal} {Phys. Rev. B}\
		}\textbf {\bibinfo {volume} {81}},\ \bibinfo {pages} {115126} (\bibinfo
		{year} {2010})}\BibitemShut {NoStop}%
	\bibitem [{\citenamefont {Peng}\ and\ \citenamefont {Lany}(2013)}]{PL13}%
	\BibitemOpen
	\bibfield  {author} {\bibinfo {author} {\bibfnamefont {H.}~\bibnamefont
			{Peng}}\ and\ \bibinfo {author} {\bibfnamefont {S.}~\bibnamefont {Lany}},\
	}\href {\doibase 10.1103/PhysRevB.87.174113} {\bibfield  {journal} {\bibinfo
			{journal} {Phys. Rev. B}\ }\textbf {\bibinfo {volume} {87}},\ \bibinfo
		{pages} {174113} (\bibinfo {year} {2013})}\BibitemShut {NoStop}%
	\bibitem [{\citenamefont {Olsen}\ and\ \citenamefont {Thygesen}(2013)}]{OT13}%
	\BibitemOpen
	\bibfield  {author} {\bibinfo {author} {\bibfnamefont {T.}~\bibnamefont
			{Olsen}}\ and\ \bibinfo {author} {\bibfnamefont {K.~S.}\ \bibnamefont
			{Thygesen}},\ }\href {\doibase 10.1103/PhysRevB.87.075111} {\bibfield
		{journal} {\bibinfo  {journal} {Phys. Rev. B}\ }\textbf {\bibinfo {volume}
			{87}},\ \bibinfo {pages} {075111} (\bibinfo {year} {2013})}\BibitemShut
	{NoStop}%
	\bibitem [{\citenamefont {Furche}(2001)}]{F01}%
	\BibitemOpen
	\bibfield  {author} {\bibinfo {author} {\bibfnamefont {F.}~\bibnamefont
			{Furche}},\ }\href@noop {} {\bibfield  {journal} {\bibinfo  {journal} {Phys.
				Rev. B}\ }\textbf {\bibinfo {volume} {64}},\ \bibinfo {pages} {195120}
		(\bibinfo {year} {2001})}\BibitemShut {NoStop}%
	\bibitem [{\citenamefont {Bracey}\ \emph {et~al.}(2009)\citenamefont {Bracey},
		\citenamefont {Ellis},\ and\ \citenamefont {Hutchings}}]{BEH09}%
	\BibitemOpen
	\bibfield  {author} {\bibinfo {author} {\bibfnamefont {C.~L.}\ \bibnamefont
			{Bracey}}, \bibinfo {author} {\bibfnamefont {P.~R.}\ \bibnamefont {Ellis}}, \
		and\ \bibinfo {author} {\bibfnamefont {G.~J.}\ \bibnamefont {Hutchings}},\
	}\href@noop {} {\bibfield  {journal} {\bibinfo  {journal} {Chemical Society
				Reviews}\ }\textbf {\bibinfo {volume} {38}},\ \bibinfo {pages} {2231}
		(\bibinfo {year} {2009})}\BibitemShut {NoStop}%
	\bibitem [{\citenamefont {Mahanty}\ and\ \citenamefont {Taylor}(1978)}]{MT78}%
	\BibitemOpen
	\bibfield  {author} {\bibinfo {author} {\bibfnamefont {J.}~\bibnamefont
			{Mahanty}}\ and\ \bibinfo {author} {\bibfnamefont {R.}~\bibnamefont
			{Taylor}},\ }\href {\doibase 10.1103/PhysRevB.17.554} {\bibfield  {journal}
		{\bibinfo  {journal} {Phys. Rev. B}\ }\textbf {\bibinfo {volume} {17}},\
		\bibinfo {pages} {554} (\bibinfo {year} {1978})}\BibitemShut {NoStop}%
	\bibitem [{\citenamefont {Fetter}\ and\ \citenamefont {Walecka}(2012)}]{FW12}%
	\BibitemOpen
	\bibfield  {author} {\bibinfo {author} {\bibfnamefont {A.~L.}\ \bibnamefont
			{Fetter}}\ and\ \bibinfo {author} {\bibfnamefont {J.~D.}\ \bibnamefont
			{Walecka}},\ }\href@noop {} {\emph {\bibinfo {title} {Quantum theory of
				many-particle systems}}}\ (\bibinfo  {publisher} {Courier Corporation},\
	\bibinfo {year} {2012})\BibitemShut {NoStop}%
	\bibitem [{\citenamefont {Colonna}\ \emph {et~al.}(2014)\citenamefont
		{Colonna}, \citenamefont {Hellgren},\ and\ \citenamefont {{de
				Gironcoli}}}]{CHG14}%
	\BibitemOpen
	\bibfield  {author} {\bibinfo {author} {\bibfnamefont {N.}~\bibnamefont
			{Colonna}}, \bibinfo {author} {\bibfnamefont {M.}~\bibnamefont {Hellgren}}, \
		and\ \bibinfo {author} {\bibfnamefont {S.}~\bibnamefont {{de Gironcoli}}},\
	}\href@noop {} {\bibfield  {journal} {\bibinfo  {journal} {Phys. Rev. B}\
		}\textbf {\bibinfo {volume} {90}},\ \bibinfo {pages} {125150} (\bibinfo
		{year} {2014})}\BibitemShut {NoStop}%
	\bibitem [{\citenamefont {Sun}\ \emph {et~al.}(2011)\citenamefont {Sun},
		\citenamefont {Marsman}, \citenamefont {Csonka}, \citenamefont {Ruzsinszky},
		\citenamefont {Hao}, \citenamefont {Kim}, \citenamefont {Kresse},\ and\
		\citenamefont {Perdew}}]{SMCRHKKP11}%
	\BibitemOpen
	\bibfield  {author} {\bibinfo {author} {\bibfnamefont {J.}~\bibnamefont
			{Sun}}, \bibinfo {author} {\bibfnamefont {M.}~\bibnamefont {Marsman}},
		\bibinfo {author} {\bibfnamefont {G.~I.}\ \bibnamefont {Csonka}}, \bibinfo
		{author} {\bibfnamefont {A.}~\bibnamefont {Ruzsinszky}}, \bibinfo {author}
		{\bibfnamefont {P.}~\bibnamefont {Hao}}, \bibinfo {author} {\bibfnamefont
			{Y.-S.}\ \bibnamefont {Kim}}, \bibinfo {author} {\bibfnamefont
			{G.}~\bibnamefont {Kresse}}, \ and\ \bibinfo {author} {\bibfnamefont {J.~P.}\
			\bibnamefont {Perdew}},\ }\href {\doibase 10.1103/PhysRevB.84.035117}
	{\bibfield  {journal} {\bibinfo  {journal} {Phys. Rev. B}\ }\textbf {\bibinfo
			{volume} {84}},\ \bibinfo {pages} {035117} (\bibinfo {year}
		{2011})}\BibitemShut {NoStop}%
	\bibitem [{\citenamefont {Bl\"{o}chl}(1994)}]{B94}%
	\BibitemOpen
	\bibfield  {author} {\bibinfo {author} {\bibfnamefont {P.~E.}\ \bibnamefont
			{Bl\"{o}chl}},\ }\href {\doibase 10.1103/PhysRevB.50.17953} {\bibfield
		{journal} {\bibinfo  {journal} {Phys. Rev. B}\ }\textbf {\bibinfo {volume}
			{50}},\ \bibinfo {pages} {17953} (\bibinfo {year} {1994})}\BibitemShut
	{NoStop}%
	\bibitem [{\citenamefont {Walter}\ \emph {et~al.}(2008)\citenamefont {Walter},
		\citenamefont {H{\"a}kkinen}, \citenamefont {Lehtovaara}, \citenamefont
		{Puska}, \citenamefont {Enkovaara}, \citenamefont {Rostgaard},\ and\
		\citenamefont {Mortensen}}]{gpaw1}%
	\BibitemOpen
	\bibfield  {author} {\bibinfo {author} {\bibfnamefont {M.}~\bibnamefont
			{Walter}}, \bibinfo {author} {\bibfnamefont {H.}~\bibnamefont
			{H{\"a}kkinen}}, \bibinfo {author} {\bibfnamefont {L.}~\bibnamefont
			{Lehtovaara}}, \bibinfo {author} {\bibfnamefont {M.}~\bibnamefont {Puska}},
		\bibinfo {author} {\bibfnamefont {J.}~\bibnamefont {Enkovaara}}, \bibinfo
		{author} {\bibfnamefont {C.}~\bibnamefont {Rostgaard}}, \ and\ \bibinfo
		{author} {\bibfnamefont {J.~J.}\ \bibnamefont {Mortensen}},\ }\href {\doibase
		10.1063/1.2943138} {\bibfield  {journal} {\bibinfo  {journal} {J. Chem.
				Phys.}\ }\textbf {\bibinfo {volume} {128}},\ \bibinfo {pages} {244101}
		(\bibinfo {year} {2008})}\BibitemShut {NoStop}%
	\bibitem [{\citenamefont {Mortensen}\ \emph {et~al.}(2005)\citenamefont
		{Mortensen}, \citenamefont {Hansen},\ and\ \citenamefont {Jacobsen}}]{gpaw2}%
	\BibitemOpen
	\bibfield  {author} {\bibinfo {author} {\bibfnamefont {J.~J.}\ \bibnamefont
			{Mortensen}}, \bibinfo {author} {\bibfnamefont {L.~B.}\ \bibnamefont
			{Hansen}}, \ and\ \bibinfo {author} {\bibfnamefont {K.~W.}\ \bibnamefont
			{Jacobsen}},\ }\href {\doibase 10.1103/PhysRevB.71.035109} {\bibfield
		{journal} {\bibinfo  {journal} {Phys. Rev. B}\ }\textbf {\bibinfo {volume}
			{71}},\ \bibinfo {pages} {035109} (\bibinfo {year} {2005})}\BibitemShut
	{NoStop}%
	\bibitem [{\citenamefont {Bahn}\ and\ \citenamefont {Jacobsen}(2002)}]{ase}%
	\BibitemOpen
	\bibfield  {author} {\bibinfo {author} {\bibfnamefont {S.~R.}\ \bibnamefont
			{Bahn}}\ and\ \bibinfo {author} {\bibfnamefont {K.~W.}\ \bibnamefont
			{Jacobsen}},\ }\href {\doibase 10.1109/5992.998641} {\bibfield  {journal}
		{\bibinfo  {journal} {Comput. Sci. Eng.}\ }\textbf {\bibinfo {volume} {4}},\
		\bibinfo {pages} {56} (\bibinfo {year} {2002})}\BibitemShut {NoStop}%
	\bibitem [{\citenamefont {Hafner}(2008)}]{VASP}%
	\BibitemOpen
	\bibfield  {author} {\bibinfo {author} {\bibfnamefont {J.}~\bibnamefont
			{Hafner}},\ }\href@noop {} {\bibfield  {journal} {\bibinfo  {journal} {J.
				Comput. Chem.}\ }\textbf {\bibinfo {volume} {29}},\ \bibinfo {pages} {2044}
		(\bibinfo {year} {2008})}\BibitemShut {NoStop}%
	\bibitem [{\citenamefont {Kresse}\ and\ \citenamefont
		{Furthm\"uller}(1996)}]{KF96}%
	\BibitemOpen
	\bibfield  {author} {\bibinfo {author} {\bibfnamefont {G.}~\bibnamefont
			{Kresse}}\ and\ \bibinfo {author} {\bibfnamefont {J.}~\bibnamefont
			{Furthm\"uller}},\ }\href {\doibase 10.1103/PhysRevB.54.11169} {\bibfield
		{journal} {\bibinfo  {journal} {Phys. Rev. B}\ }\textbf {\bibinfo {volume}
			{54}},\ \bibinfo {pages} {11169} (\bibinfo {year} {1996})}\BibitemShut
	{NoStop}%
	\bibitem [{\citenamefont {Kresse}\ and\ \citenamefont {Joubert}(1999)}]{KJ99}%
	\BibitemOpen
	\bibfield  {author} {\bibinfo {author} {\bibfnamefont {G.}~\bibnamefont
			{Kresse}}\ and\ \bibinfo {author} {\bibfnamefont {D.}~\bibnamefont
			{Joubert}},\ }\href@noop {} {\bibfield  {journal} {\bibinfo  {journal} {Phys.
				Rev. B}\ }\textbf {\bibinfo {volume} {59}},\ \bibinfo {pages} {1758}
		(\bibinfo {year} {1999})}\BibitemShut {NoStop}%
	\bibitem [{\citenamefont {Togo}\ and\ \citenamefont {Tanaka}(2015)}]{TT15}%
	\BibitemOpen
	\bibfield  {author} {\bibinfo {author} {\bibfnamefont {A.}~\bibnamefont
			{Togo}}\ and\ \bibinfo {author} {\bibfnamefont {I.}~\bibnamefont {Tanaka}},\
	}\href@noop {} {\bibfield  {journal} {\bibinfo  {journal} {Scr. Mater.}\
		}\textbf {\bibinfo {volume} {108}},\ \bibinfo {pages} {1} (\bibinfo {year}
		{2015})}\BibitemShut {NoStop}%
	\bibitem [{\citenamefont {Birch}(1947)}]{B71}%
	\BibitemOpen
	\bibfield  {author} {\bibinfo {author} {\bibfnamefont {F.}~\bibnamefont
			{Birch}},\ }\href {\doibase 10.1103/PhysRev.71.809} {\bibfield  {journal}
		{\bibinfo  {journal} {Phys. Rev.}\ }\textbf {\bibinfo {volume} {71}},\
		\bibinfo {pages} {809} (\bibinfo {year} {1947})}\BibitemShut {NoStop}%
	\bibitem [{\citenamefont {Bates}\ \emph {et~al.}(2016)\citenamefont {Bates},
		\citenamefont {Laricchia},\ and\ \citenamefont {Ruzsinszky}}]{BLR16}%
	\BibitemOpen
	\bibfield  {author} {\bibinfo {author} {\bibfnamefont {J.~E.}\ \bibnamefont
			{Bates}}, \bibinfo {author} {\bibfnamefont {S.}~\bibnamefont {Laricchia}}, \
		and\ \bibinfo {author} {\bibfnamefont {A.}~\bibnamefont {Ruzsinszky}},\
	}\href@noop {} {\bibfield  {journal} {\bibinfo  {journal} {Phys. Rev. B}\
		}\textbf {\bibinfo {volume} {93}},\ \bibinfo {pages} {045119} (\bibinfo
		{year} {2016})}\BibitemShut {NoStop}%
	\bibitem [{\citenamefont {Gr{\"u}neis}\ \emph {et~al.}(2009)\citenamefont
		{Gr{\"u}neis}, \citenamefont {Marsman}, \citenamefont {Harl}, \citenamefont
		{Schimka},\ and\ \citenamefont {Kresse}}]{GMHSK09}%
	\BibitemOpen
	\bibfield  {author} {\bibinfo {author} {\bibfnamefont {A.}~\bibnamefont
			{Gr{\"u}neis}}, \bibinfo {author} {\bibfnamefont {M.}~\bibnamefont
			{Marsman}}, \bibinfo {author} {\bibfnamefont {J.}~\bibnamefont {Harl}},
		\bibinfo {author} {\bibfnamefont {L.}~\bibnamefont {Schimka}}, \ and\
		\bibinfo {author} {\bibfnamefont {G.}~\bibnamefont {Kresse}},\ }\href@noop {}
	{\bibfield  {journal} {\bibinfo  {journal} {The Journal of chemical physics}\
		}\textbf {\bibinfo {volume} {131}},\ \bibinfo {pages} {154115} (\bibinfo
		{year} {2009})}\BibitemShut {NoStop}%
	\bibitem [{\citenamefont {Janthon}\ \emph {et~al.}(2014)\citenamefont
		{Janthon}, \citenamefont {Luo}, \citenamefont {Kozlov}, \citenamefont
		{Vines}, \citenamefont {Limtrakul}, \citenamefont {Truhlar},\ and\
		\citenamefont {Illas}}]{JLKVLTI14}%
	\BibitemOpen
	\bibfield  {author} {\bibinfo {author} {\bibfnamefont {P.}~\bibnamefont
			{Janthon}}, \bibinfo {author} {\bibfnamefont {S.}~\bibnamefont {Luo}},
		\bibinfo {author} {\bibfnamefont {S.~M.}\ \bibnamefont {Kozlov}}, \bibinfo
		{author} {\bibfnamefont {F.}~\bibnamefont {Vines}}, \bibinfo {author}
		{\bibfnamefont {J.}~\bibnamefont {Limtrakul}}, \bibinfo {author}
		{\bibfnamefont {D.~G.}\ \bibnamefont {Truhlar}}, \ and\ \bibinfo {author}
		{\bibfnamefont {F.}~\bibnamefont {Illas}},\ }\href@noop {} {\bibfield
		{journal} {\bibinfo  {journal} {Journal of chemical theory and computation}\
		}\textbf {\bibinfo {volume} {10}},\ \bibinfo {pages} {3832} (\bibinfo {year}
		{2014})}\BibitemShut {NoStop}%
	\bibitem [{\citenamefont {Huntington}(1958)}]{H58}%
	\BibitemOpen
	\bibfield  {author} {\bibinfo {author} {\bibfnamefont {H.~B.}\ \bibnamefont
			{Huntington}},\ }\href@noop {} {\emph {\bibinfo {title} {Solid state
				physics}}},\ Vol.~\bibinfo {volume} {7}\ (\bibinfo  {publisher} {Elsevier},\
	\bibinfo {year} {1958})\ pp.\ \bibinfo {pages} {213--351}\BibitemShut
	{NoStop}%
	\bibitem [{\citenamefont {Cardarelli}(2018)}]{C18}%
	\BibitemOpen
	\bibfield  {author} {\bibinfo {author} {\bibfnamefont {F.}~\bibnamefont
			{Cardarelli}},\ }\href {\doibase 10.1007/978-3-319-38925-7_4} {\emph
		{\bibinfo {title} {Materials Handbook: A Concise Desktop Reference}}}\
	(\bibinfo  {publisher} {Springer International Publishing},\ \bibinfo
	{address} {Cham},\ \bibinfo {year} {2018})\ pp.\ \bibinfo {pages}
	{317--695}\BibitemShut {NoStop}%
	\bibitem [{\citenamefont {O'Hara}\ and\ \citenamefont {Marshall}(1971)}]{OM71}%
	\BibitemOpen
	\bibfield  {author} {\bibinfo {author} {\bibfnamefont {S.~G.}\ \bibnamefont
			{O'Hara}}\ and\ \bibinfo {author} {\bibfnamefont {B.~J.}\ \bibnamefont
			{Marshall}},\ }\href {\doibase 10.1103/PhysRevB.3.4002} {\bibfield  {journal}
		{\bibinfo  {journal} {Phys. Rev. B}\ }\textbf {\bibinfo {volume} {3}},\
		\bibinfo {pages} {4002} (\bibinfo {year} {1971})}\BibitemShut {NoStop}%
	\bibitem [{\citenamefont {Teodosiu}(2013)}]{T13}%
	\BibitemOpen
	\bibfield  {author} {\bibinfo {author} {\bibfnamefont {C.}~\bibnamefont
			{Teodosiu}},\ }\href@noop {} {\emph {\bibinfo {title} {Elastic models of
				crystal defects}}}\ (\bibinfo  {publisher} {Springer Science \& Business
		Media},\ \bibinfo {year} {2013})\BibitemShut {NoStop}%
	\bibitem [{\citenamefont {Simmons}(1965)}]{S65}%
	\BibitemOpen
	\bibfield  {author} {\bibinfo {author} {\bibfnamefont {G.}~\bibnamefont
			{Simmons}},\ }\href@noop {} {\emph {\bibinfo {title} {Single crystal elastic
				constants and calculated aggregate properties}}},\ \bibinfo {type} {Tech.
		Rep.}\ (\bibinfo  {institution} {SOUTHERN METHODIST UNIV DALLAS TEX},\
	\bibinfo {year} {1965})\BibitemShut {NoStop}%
	\bibitem [{\citenamefont {Perdew}\ \emph {et~al.}(2009)\citenamefont {Perdew},
		\citenamefont {Ruzsinszky}, \citenamefont {Csonka}, \citenamefont
		{Constantin},\ and\ \citenamefont {Sun}}]{PRCCS09}%
	\BibitemOpen
	\bibfield  {author} {\bibinfo {author} {\bibfnamefont {J.~P.}\ \bibnamefont
			{Perdew}}, \bibinfo {author} {\bibfnamefont {A.}~\bibnamefont {Ruzsinszky}},
		\bibinfo {author} {\bibfnamefont {G.~I.}\ \bibnamefont {Csonka}}, \bibinfo
		{author} {\bibfnamefont {L.~A.}\ \bibnamefont {Constantin}}, \ and\ \bibinfo
		{author} {\bibfnamefont {J.}~\bibnamefont {Sun}},\ }\href@noop {} {\bibfield
		{journal} {\bibinfo  {journal} {Physical Review Letters}\ }\textbf {\bibinfo
			{volume} {103}},\ \bibinfo {pages} {026403} (\bibinfo {year}
		{2009})}\BibitemShut {NoStop}%
	\bibitem [{\citenamefont {Nguyen}\ \emph {et~al.}(2014)\citenamefont {Nguyen},
		\citenamefont {Colonna},\ and\ \citenamefont {de~Gironcoli}}]{NCG14}%
	\BibitemOpen
	\bibfield  {author} {\bibinfo {author} {\bibfnamefont {N.~L.}\ \bibnamefont
			{Nguyen}}, \bibinfo {author} {\bibfnamefont {N.}~\bibnamefont {Colonna}}, \
		and\ \bibinfo {author} {\bibfnamefont {S.}~\bibnamefont {de~Gironcoli}},\
	}\href {\doibase 10.1103/PhysRevB.90.045138} {\bibfield  {journal} {\bibinfo
			{journal} {Phys. Rev. B}\ }\textbf {\bibinfo {volume} {90}},\ \bibinfo
		{pages} {045138} (\bibinfo {year} {2014})}\BibitemShut {NoStop}%
	\bibitem [{\citenamefont {Jauho}\ \emph {et~al.}(2015)\citenamefont {Jauho},
		\citenamefont {Olsen}, \citenamefont {Bligaard},\ and\ \citenamefont
		{Thygesen}}]{JOBT15}%
	\BibitemOpen
	\bibfield  {author} {\bibinfo {author} {\bibfnamefont {T.~S.}\ \bibnamefont
			{Jauho}}, \bibinfo {author} {\bibfnamefont {T.}~\bibnamefont {Olsen}},
		\bibinfo {author} {\bibfnamefont {T.}~\bibnamefont {Bligaard}}, \ and\
		\bibinfo {author} {\bibfnamefont {K.~S.}\ \bibnamefont {Thygesen}},\
	}\href@noop {} {\bibfield  {journal} {\bibinfo  {journal} {Phys. Rev. B}\
		}\textbf {\bibinfo {volume} {92}},\ \bibinfo {pages} {115140} (\bibinfo
		{year} {2015})}\BibitemShut {NoStop}%
	\bibitem [{\citenamefont {Kittel}\ \emph {et~al.}(1976)\citenamefont {Kittel},
		\citenamefont {McEuen},\ and\ \citenamefont {McEuen}}]{KMM76}%
	\BibitemOpen
	\bibfield  {author} {\bibinfo {author} {\bibfnamefont {C.}~\bibnamefont
			{Kittel}}, \bibinfo {author} {\bibfnamefont {P.}~\bibnamefont {McEuen}}, \
		and\ \bibinfo {author} {\bibfnamefont {P.}~\bibnamefont {McEuen}},\
	}\href@noop {} {\emph {\bibinfo {title} {Introduction to solid state
				physics}}},\ Vol.~\bibinfo {volume} {8}\ (\bibinfo  {publisher} {Wiley New
		York},\ \bibinfo {year} {1976})\BibitemShut {NoStop}%
\end{thebibliography}
\end{document}